\documentclass[a4paper,11pt,english]{article}\pdfoutput=1\usepackage{jheppub}
\usepackage[utf8]{inputenc} 
\usepackage{physics}
\usepackage{babel}

\usepackage{graphicx,color,overpic,mathtools}

\usepackage{cancel}
\usepackage[normalem]{ulem}

\begin{document}

\title{Reissner-Nordström geometry counterpart in semiclassical gravity}

\author[1]{Julio Arrechea,}
\author[1]{Carlos Barcel\'o,} 
\affiliation[1]{Instituto de Astrof\'isica de Andaluc\'ia (IAA-CSIC),
Glorieta de la Astronom\'ia, 18008 Granada, Spain}
\author[2]{Ra\'ul Carballo-Rubio}
\affiliation[2]{Florida Space Institute, University of Central Florida, 12354 Research Parkway, Partnership 1, Orlando, FL, USA}
\author[3,4]{and Luis J. Garay} 
\affiliation[3]{Departamento de F\'{\i}sica Te\'orica and IPARCOS, Universidad Complutense de Madrid, 28040 Madrid, Spain}  
\affiliation[4]{Instituto de Estructura de la Materia (IEM-CSIC), Serrano 121, 28006 Madrid, Spain}
\emailAdd{arrechea@iaa.es}
\emailAdd{carlos@iaa.es}
\emailAdd{raul.carballorubio@ucf.edu}
\emailAdd{luisj.garay@ucm.es}

\abstract{We compute the Renormalized Stress-Energy Tensor (RSET) of a massless minimally coupled scalar field in the Regularized Polyakov approximation, as well as its backreaction, on the classical Reissner-Nordström spacetime. The complete set of solutions of the semiclassical self-consistent equations is obtained and compared with the classical counterparts. The semiclassical Reissner-Nordström family involves three kinds of geometries that depend on the charge-to-mass ratio of the spacetime. In the under-charged regime, the geometry has its external horizon replaced by a wormhole neck that leads to a singular asymptotic region at finite proper distance. The over-charged regime reveals a naked singularity coated by a cloud of (infinite) mass coming from the quantized field. In between both behaviours there is a separatrix solution reminiscent of the extremal black hole classical geometry. As the RSET over an extremal horizon is finite, the semiclassical backreaction does not get rid of the horizon. Nonetheless, we show that the resulting horizon is singular.}

\maketitle

\section{Introduction}

In the search for the first quantum gravity effects accessible to our current understanding, the study of black-hole spacetimes subject to semiclassical effects have become a successful test bench. Quantum field theory in curved spacetimes has transformed our understanding of the nature of black holes. Landmark predictions of quantum field theory in curved spacetimes include black hole evaporation via the emission of Hawking quanta~\cite{Hawking1975}, the Unruh effect~\cite{Unruh1976}, and cosmological particle creation~\cite{Parker1968}. Furthermore, the interest raised by the black hole evaporation paradigm~\cite{Hawking1974} has been one of the main fuel in developing frameworks beyond general relativity. It has also been one of the main incentives for the analysis of different deviations from general relativistic black holes (see e.g.~\cite{Mathur2005, Mazur2001, Barcelo2009}), together with the possibility of detecting these deviations observationally~\cite{Carballo-Rubio:2018jzw,Cardoso:2019rvt}.

Semiclassical gravity emerges as an implementation of quantum field theory in curved spacetimes that treats spacetime classically while the material content giving rise to such spacetime is of mixed, quantum and classical, nature. Thus, finding self-consistent semiclassical solutions amounts to solving the Einstein equations using the Renormalized Stress Energy Tensor (RSET) of quantum fields as an additional source of gravity. Before finding solutions sourced by specific classical matter, it is convenient to understand the structural characteristics of vacuum semiclassical solutions.

Previous analyses~\cite{Fabbrietal2005, Berthiere2017, Ho2017, Arrechea2019} have tackled the problem of finding the semiclassical counterpart to the Schwarzschild spacetime, that is, the vacuum solutions of semiclassical general relativity. To that end, the RSET of a massless scalar field (the simplest model to analyze) was taken in the Boulware vacuum, the one vacuum state compatible with both static and asymptotically flat geometries. The Boulware vacuum is singular at the Schwarzschild horizon, indicating the presence of strong backreaction effects there. Indeed, the resulting self-consistent spacetimes---incorporating backreaction---have no horizon and are, instead of black holes, asymmetric wormholes with one singular end. The resulting wormholes have one asymptotically flat end, whereas the other end develops a null singularity located at a finite proper distance from the neck of the wormhole. As opposed to the Schwarwschild geometry, its semiclassical counterpart exhibits a singularity which is not covered by any event horizon (for a discussion of potential lessons to be learned from this fact see~\cite{Arrechea2020}). 

In this work, we extend the analysis in~\cite{Arrechea2019} considering here the electrovacuum solutions of semiclassical general relativity. Again, for the quantum sector, we compute the vacuum polarization of a single massless scalar field minimally coupled to curvature. For the RSET we use the so called Regularized Polyakov (RP-RSET) approximation described in detail in~\cite{Arrechea2019}. At the beginning of the next section we will discuss the reasons behind the use of this approximation and the associated benefits and limitations. Armed with this RSET, in this work we characterize the complete set of semiclassical electrovacuum solutions, which can be separated into three subclasses attending to whether the geometry is under-charged, over-charged or balanced. We numerically compute solutions belonging to each of the subclasses, and we also find analytical approximations to the solutions in certain spacetime regions.

The Reissner-Nordström geometry has been used in several analyses of quantum-induced phenomena. One notorious characteristic is that the classical sub-extremal Reisner-Nordström black hole shows a double horizon structure, with a future outer horizon and a future inner horizon (we are using the notation first introduced in \cite{Hayward1993}). This structure is similar to that exhibited by Kerr black holes, while preserving the benevolence of spherical symmetry. That is why the analysis of Reissner-Nordström geometries is also used as a proxy to understanding the more complicated Kerr case. The presence of a second internal horizon adds some special features to these geometries. For example, different analyses show that, as opposed to the outer horizon, the inner horizon is highly unstable under both classical and quantum perturbations~\cite{Ori1991,Balbinot1993}. Our analysis here offers an additional perspective into these stability-related aspects: As we will see, the semiclassical corrections eliminate the outer horizon in such a way that the system never enters (at least in vacuum) a regime which explores the physics of an inner horizon. This could be taken as a suggestion that horizons are only present during transient regimes but are absent in genuinely static configurations.

Another intereresting aspect of the Reissner-Norsdtröm family is the existence of an extremal solution, in which inner and outer horizons are spatially coincident. In the Boulware vacuum, the value at the RSET at the outer horizon diverges, but not if the horizon is extremal~\cite{Anderson1995, Fagnocchi2005, Farese2005}. This can be taken as an indication that extremal black holes are stable under semiclassical effects. Perhaps surprisingly, this is not what we find. Semiclassical corrections displace the extremal horizon from its original position and transforms it into a curvature singularity, unveiling a narrow region beyond this singular horizon where semiclassical corrections become non-perturbative. This result exemplifies a situation where calculating the RSET over a fixed background geometry suggest that semiclassical corrections act perturbatively around the horizon, but the geometry incorporating back-reaction self-consistently develops singularities. In view of this, we point out the existence of regimes in which predictions based only on fixed-background computations should be taken with due care.
   
The previous results seem to indicate a strong tension between quantized fields living in the Boulware state and event horizons of any kind. It has been argued that horizons with a higher multiplicity may be compatible with a finite RSET \cite{Anderson1998}. As these geometries cannot be attained without the introduction of additional matter fields, we are leaving them outside our discussion. On the other hand, by considering the fields to be in the Hartle-Hawking vacuum state, the local structure of the outer horizon of the Reissner-Nordström geometry can be maintained. This is accomplished by the introduction of fluxes of energy that diverge at the horizon, thus cancelling the Boulware divergence. The inner portion of the geometry becomes unveiled and allows to explore how the Polyakov RSET backreacts near the inner horizon. A detailed analysis on this topic will be presented elsewhere.

Finally, let us mention that in order to analyze the structure of semiclassical charged compact objects, these must be matched at their surface with the semiclassical Reissner-Nordström geometry here depicted. The repulsion exerted by the electromagnetic field allows charged spheres of fluid to reach arbitrarily high compactness \cite{deFelice1999}. Hence, these configurations could serve as models in which the effects of quantized fields in highly compact scenarios could be analyzed.

\subsection{Summary of the results and structure of the paper}

In the same spirit as in~\cite{Arrechea2019}, here we find the semiclassical Reissner-Nordström counterparts. The presence of the electromagnetic field gives rise to a broader spectrum of geometries.

In our analysis we find three families of solutions. The first one, corresponding to spacetimes where the electromagnetic charge is below the mass, has a wormhole structure similar to that of the semiclassical Schwarzschild counterpart, and connects to that solution in the vanishing charge limit. As charge is increased, electromagnetic repulsion makes the wormhole neck shrink (as the classical horizon would do), but the structure of the asymptotic singularity at the other side of the neck remains unmodified, as the effects of charge decay with distance. 

The second family where charge surpasses mass resembles naked singularities, but this time located at finite distance in the $r$ coordinate. Contrary to the previous case where the RP-RSET backreacts to make the geometry devoid of horizons, the super-charged case corresponds to a naked singularity in the classical regime. From this situation we learn how vacuum polarization backreacts on an already singular geometry, which turns out to result in an increased strength of the singularity. The dominant contribution at short distances comes from a runaway of vacuum polarization, stimulated, in part, by the blowing up of the Polyakov RSET as we approach the center of the geometry. The electromagnetic charge, however, retains its ``undressed" value and does not contribute to increasing the strength of the singularity. In situations where the RP-RSET is sufficiently suppressed, the dominant divergence contribution to the mass comes from terms proportional to the charge rather than those coming from vacuum polarization.

In between both cases there is a separatrix solution, or “quasi-extremal” geometry, with a degenerate horizon where vacuum polarization is finite. The existence of the semiclassical extremal black hole has been subject of debate in the community \cite{Trivedi1992, Anderson2000,  Lowe2000,  Matyjasek2001}. It has been shown, through different approximations to the RSET, that its components are well-behaved at the extremal horizon \cite{Anderson1995, Fagnocchi2005, Farese2005}. This suggests that backreaction will modify mildly the horizon structure of the extremal black hole, but without destroying it altogether. With these considerations in mind we present a self-consistent semiclassical “quasi-extremal” black hole geometry in 3 + 1 dimensions and analyze some of its properties. However, we observe that this “quasi-extremal” geometry develops a non-analyticity at the horizon, in the form of a cusp in the metric functions, and that this behaviour generates a curvature singularity. This curvature singularity is not visible in curvature scalar invariants. Instead, it appears when the Riemann curvature tensor is contracted with a tetrad field parallel transported along an infalling geodesic trajectory. Backreaction on the extremal black hole is, in this sense, more benign than for spacetimes whose horizons have the same local form as the Schwarzschild horizon. The horizon characteristic is preserved, though the effect of vacuum polarization makes this horizon singular. This result strongly suggests the incompatibility between the Boulware vacuum state and regular horizons of any sort.

The paper is organised as follows. Section \ref{Sec:prelim} reviews the classical Reissner-Nordström family and some of its properties, alongside a discussion of the Regularized Polyakov RSET. Section \ref{Sec:Self-consistent} sets up the analysis for the self-consistent electro-vacuum semiclassical equations. Sections \ref{Sec:Wormhole}, \ref{Sec:Naked} and \ref{Sec:Extremal} analyze sub-charged, super-charged and quasi-extremal regimes, respectively. Finally, section \ref{Sec:Conclusions} is reserved for discussion and conclusions.
 
\section{Preliminaries}
\label{Sec:prelim}
\subsection{The classical electro-vacuum solution}

We consider the most general static and spherically symmetric line element
\begin{equation}\label{Eq:LineElement}
ds^{2}=-e^{2\phi(r)}dt^{2}+\frac{1}{1-C(r)}dr^{2}+r^{2}d\Omega^{2},
\end{equation}
where $e^{2\phi}$ is the redshift function encoding the redshift suffered by escaping lightrays, and $C$ is the compactness function, which can be written as $C(r)=2m(r)/r$ where $m(r)$ is the Misner-Sharp mass contained inside spheres of radius $r$ \cite{Misner1964,Hernandez1966,Hayward1994}. In vacuum, the time component of the geometry equals the inverse of the radial one, but this relation no longer holds in the presence of matter (note that the electromagnetic stress-energy tensor (SET) constitutes an exception). The Reissner-Nordström spacetime is the geometry that results from solving the Einstein-Maxwell equations under the assumptions of staticity and spherical symmetry. From the Maxwell equations we obtain the following form for the electromagnetic SET,
\begin{equation}\label{Eq:Set}
T_{\mu}^{\nu}=\text{diag}(-1,-1,1,1)\frac{Q^{2}}{8\pi r^{4}}.
\end{equation}
Here, $Q^{2}=Q_{\text{e}}^{2}+Q_{\text{m}}^{2}$, where $Q_{\text{e}}$ and $Q_{\text{m}}$ denote the electric and magnetic charges, respectively. Greek letters denote spacetime indexes. The $tt$ and $rr$ components of the classical field equations are
\begin{align}
C+rC'
&
=\frac{Q^{2}}{r^{2}},\nonumber\\
-2r\psi+C(1+2r\psi)
&
=\frac{Q^{2}}{r^{2}},\label{Eq:Fieldeqsclas}
\end{align}
where the $'$ denotes derivatives with respect to the $r$ coordinate and $\psi\equiv\phi'$. Solving equations \eqref{Eq:Fieldeqsclas} yields the Reissner-Nordström geometry
\begin{equation}
ds^{2}=-\left(1-\frac{2M}{r}+\frac{Q^{2}}{r^{2}}\right)dt^{2}+\left(1-\frac{2M}{r}+\frac{Q^{2}}{r^{2}}\right)^{-1}dr^{2}+r^{2}d\Omega^{2}.
\end{equation}
This solution shows several unique features due to the presence of charge. The zeroes of the redshift function determine the location of its two horizons 
\begin{equation}\label{Eq:RNHor}
r_{\pm}=M\pm\sqrt{M^{2}-Q^{2}}.
\end{equation}
Therefore, depending on the charge-to-mass ratio of the geometry, it can exhibit two, one, or no horizons whatsoever. This splits the Reissner-Nordström family into three cathegories: 
\begin{itemize}
\item Sub-extremal ($Q<M$), which shows a timelike singularity covered by outer and inner horizons at $r_{-}$ and $r_{+}$ respectively. The Schwarzschild black hole is the particular case where $Q=0$.
\item Super-extremal ($Q>M$), where the value of the charge surpassing that of the mass is understood as if some charged negative mass had entered the geometry. Ultimately, this gives rise to a naked singularity. 
\item Extremal black holes ($Q=M$), for which outer and inner horizons become coincident, forming an extremal horizon at the end of an infinite neck.
 \end{itemize}

\subsection{On the different approximations to the RSET}
\label{Subsec:Different RSET}

Whereas in flat spacetime infinite vacuum energies can be subtracted, this procedure becomes intricate in curved spacetimes where, in order to account for the genuine contribution of vacuum energy to the spacetime curvature, the field stress-energy tensor undergoes a renormalization procedure \cite{Birrell1984}. Ideally, one would hope to have a single exact RSET, but in practice the situation is much more complicated. On the one hand, we have ambiguities in the renormalization procedure~\cite{Wald1995}. On the other hand, we have different approximation schemes which can be well-suited to different aspects of the problem.

Concerning the various approximation schemes, there is a range from basic to more sophisticated: stress-energy tensors that only inform about the $s$-mode of the fields, with or without backscattering \cite{DaviesFulling1977, Fabbri2005}; stress-energy tensors \cite{Frolov1987, Brown1986} accounting for local contributions to curvature; and more elaborate expressions adapted to scalar fields of arbitrary mass and coupling in static spacetimes, where arbitrarily high multipoles are considered \cite{Andersonetal1995, Popov2003}. Quantum corrections for the Reissner-Nordström spacetime have been calculated in some of these approximations \cite{Huang1992, Andersonetal1995, Taylor1999} and also making use of the celebrated trace anomaly \cite{Anderson2007, Abedi2017}.

It is worth mentioning the pursue for a complete calculation of the RSET by Anderson et al. \cite{Andersonetal1995}. Their RSET can be split into two independently conserved numerical and analytical parts. The analytical portion includes both high-frequency, local contributions and low-frequency, state-dependent terms \cite{Popov2003}. Ideally, one would rather solve the full backreaction equations equipped with this analytical approximation to the RSET. The complexity of the system of equations involved makes this task difficult, although solutions describing symmetric wormhole spacetimes \cite{Hochberg1977} have been found. Moreover, this analytical approximation to the RSET has been computed over the Reissner-Nordström spacetime \cite{Andersonetal1995}, showing a pathological behaviour at the horizon that disappears after the addition of the numerical part.

It is also important to stress that most of the aforementioned renormalized stress-energy tensor (RSET) approximations share the presence of higher-derivative terms that hinder a full self-consistent treatment of the semiclassical equations, where spacetime geometry and material sources are computed simultaneously. In consequence, vacuum polarization is either computed on top of a fixed background, or its backreaction is considered but not in a complete self-consistent way \cite{Parker1993}. Given the complexity of these approaches to implement and interpret self-consistent solutions, here we follow a different strategy. In 1 + 1 dimensions, the renormalized stress-energy tensor of a massless scalar field acquires a simpler form, obtainable via the point-splitting regularization method \cite{DaviesFulling1977}. After renormalization, the RSET is then transformed into a (3 + 1)-dimensional quantity by means of the Polyakov approximation \cite{Polyakov1981}. In this process we lose information about quantum fluctuations not living in the s-wave sector of the 4-dimensional spacetime. Conservation demands adding a $1/4\pi r^{2}$ multiplicative factor (with $r$ the coordinate denoting the areal radius) to the RSET components. The resulting tensor has vanishing angular components and diverges in the $r\to0$ limit, but has the strong advantage of preserving second-order field equations, making the resolution of the semiclassical equations not excessively complex. It is in $1+1$ dimensions, in the context of dilatonic gravity, where backreaction studies have flourished \cite{Hayward1995, Buric1999, Zaslavskii2002}, serving as test grounds for the physics that the full $3+1$ theory could exhibit. 

The next, natural increase in complexity consists on including backscattering of the $s$-wave field modes in the RSET \cite{Fabbri2005}. Unfortunately, we lack an approximation that incorporates backscattering and that is regular at $r=0$. Motivated by the search of RSET approximations that are both regular at the center of spherically symmetric spacetimes and contains up to second-order derivatives of the metric functions, we follow \cite{Parentani1994, Ayal1997, Arrechea2019} by adopting the Regularized Polyakov approximation, where the Polyakov multiplicative factor is modified by the introduction of a regulator that acts as a cutoff to the magnitude of the Regularized Polyakov RSET (RP-RSET) components. The RP-RSET is conserved, finite at the radial origin and has tangential pressures, features shared by the stress-energy tensor found by Anderson et al. \cite{Andersonetal1995}. Note that these convenient properties are accomplished simultaneously due to staticity. In dynamical scenarios \cite{Parentani1994, Ayal1997} the introduction of the regulator breaks conservation in such a way that cannot be compensated solely by adding angular components. Thus, in such cases, the regularized tensor no longes satisfies the covariant conservation condition. However, here we will only deal with static configurations for which the Regularized Polyakov RSET is appropriate.

\subsection{The Polyakov RSET and its regularization}
\label{Subsec:RPRSET}

Having reviewed the classical Reissner-Nordström family and some of its features, we now turn to the realm of semiclassical gravity. The semiclassical Einstein equations
\begin{equation}\label{Eq:Einstein}
G_{\mu\nu}=8\pi\left(T_{\mu\nu}+\hbar\langle\hat{T}_{\mu\nu}\rangle\right)
\end{equation}
can, in some particularly simple scenarios, be solved at the self-consistent level, where spacetime geometry and matter contributions are determined simultaneously.
For this purpose, the quantum matter contribution $\langle\hat{T}_{\mu\nu}\rangle$ is constructed by means of the Polyakov approximation. The Polyakov approximation involves a dimensional reduction to a $(1+1)$ dimensional manifold described by the non-angular sector of the metric \eqref{Eq:LineElement}. Owing to the fact that the wave equation for a scalar field propagating on top of a $(1+1)$ spacetime is conformally invariant, an exact expression for the RSET is obtained after point-splitting renormalization \cite{DaviesFulling1977}, its components being
\begin{align}\label{eq:RSETcomponents}
        \langle{\hat{T}_{rr}}\rangle^{\rm{P}2}=
        &
        -\frac{l_{\rm P}^{2}\psi^{2}}{2}+\langle\text{SDT}\rangle,\quad \quad \langle{\hat{T}_{tr}}\rangle^{\rm{P}2}=\langle{\hat{T}_{rt}}\rangle^{\rm{P}2}=0,\nonumber \\
        \langle{\hat{T}_{tt}}\rangle^{\rm{P}2}=
        &
        \frac{l_{\rm P}^{2}e^{2\phi}}{2}\left[2\psi'(1-C)+\psi^{2}(1-C)-\psi C'\right]+\langle\text{SDT}\rangle.
\end{align}
Here, $l_{\rm P}=\hbar/\sqrt{12\pi}$ and $\langle\text{SDT}\rangle$ denotes the state-dependent part of the Polyakov RSET, vanishing for the Boulware vacuum. The components \eqref{eq:RSETcomponents} are then used in order to construct the temporal and radial sector of the $(3+1)$ Polyakov RSET in the following way:
\begin{equation}\label{eq:DimTransf}
        \langle{\hat{T}_{\mu\nu}}\rangle^{\rm{P}}=\frac{1}{4\pi r^{2}}\delta^a_\mu\delta^b_\nu\langle{\hat{T}_{ab}}\rangle^{\rm{P}2},
\end{equation}
with latin indexes taking the $t,r$ values. The $(1+1)$ components must be divided by the surface area of the sphere to ensure $(3+1)$-dimensional conservation of the RSET. This multiplicative factor introduces a generic divergence in the components of the Polyakov RSET as these approach $r\to0$. 
Indeed, the Polyakov RSET has singular components when computed over geometries with finite curvature invariants at $r=0$. This comes in conflict with the idea that, in regular matter distributions of small compactness, the Polyakov RSET should amount to a tiny contribution throughout the whole spacetime. More explicitly, demanding that the Kretschmann scalar 
\begin{align}\label{Eq:Kretschmann}
     \mathcal{K}=\frac{4C^{2}}{r^{4}}+\frac{2C'^{2}}{r^{2}}+\frac{8\psi^{2}(1-C)^{2} }{r^{2}}
     +\left[\psi C'-2(\psi^{2}+\psi')(1-C)\right]^{2}
\end{align}
remains regular imposes the following behaviour for the metric functions
\begin{equation}\label{Eq:RegConds}
\phi(r)=\phi_{0}+\phi_{1}r^{2}+\mathcal{O}(r^{3}),\qquad C(r)= C_{1}r^{2}+\mathcal{O}(r^{3}),
\end{equation}
where $\phi_{0},\phi_{1}, C_{1}$ are arbitrary, non-zero constants. Now, it can be easily checked that, for the profiles \eqref{Eq:RegConds}, the semiclassical energy density 
\begin{equation}\label{Eq:SemiDens}
\rho_{\text{s}}=e^{-2\phi}\langle\hat{T}_{tt}\rangle^{\rm P}\propto\frac{\phi_{1}}{r^{2}}+\mathcal{O}(r^{0})
\end{equation}
diverges quadratically in $r$. This issue is aggravated by the fact that the semiclassical equations, due to their nonlinear nature, have this divergence displaced to $r=l_{\rm P}$. As a consequence, the Polyakov approximation breaks down at distances where we were not expecting singularities to arise. Strictly speaking, the Polyakov approximation cannot be trusted at small enough scales \cite{Carballo-Rubio2017}. In this work, our strategy is to regularize the Polyakov approximation to avoid these types of divergences, and to analyze the resulting set of solutions. Along the way, we will comment on which characteristics of the solutions found should be independent of our final approximation scheme and which ones might not be.

Inspired by the approach followed in \cite{Parentani1994}, we provided a regularization scheme for the Polyakov RSET  \cite{Arrechea2019}. This procedure is carried out in two steps. First, we introduce a cutoff in the non-angular sector of the Polyakov RSET by transforming the multiplicative factor
\begin{equation}\label{eq:DistPol}
        \langle{\hat{T}_{ab}}\rangle^{\rm DP}\to \frac{4\pi r^{2}}{4\pi \left(r^{2}+\alpha l_{\rm P}^{2}\right)}\langle{\hat{T}_{ab}}\rangle^{\rm P},
\end{equation}
where the super-index $\rm{DP}$ stands for Distorted Polyakov. Taking $\alpha>0$ ensures the regularity of $\langle{\hat{T}_{ab}}\rangle^{\rm DP}$ at the radial origin. The second step consists in adding a Compensatory piece to the Distorted Polyakov RSET so the sum of both terms gives a covariantly conserved tensor. The RP-RSET is then defined as
\begin{equation}
    T^{\rm RP}_{\mu\nu}\equiv T^{\rm DP}_{\mu\nu}+T^{\rm C}_{\mu\nu},
\end{equation}
where the components of the compensatory tensor are assumed to be angular only for simplicity, and come from algebraically solving
\begin{equation}\label{eq:conservation}
    \nabla^{\mu}T^{\rm RP}_{\mu r}=0
\end{equation}
Neither the choice of regulating factor \eqref{eq:DistPol}, nor the assumption of $\langle \hat{T}_{\mu\nu}\rangle^{\rm C}$ having only angular components, are unique. Our choice is based on an attempt to modify the Polyakov RSET in the mildest way, while achieving the desired regularity properties. The angular components of the RP-RSET have the form
\begin{equation}
    \langle \hat{T}_{\theta\theta}\rangle^{\rm RP}=\frac{\langle \hat{T}_{\varphi\varphi}\rangle^{\rm RP}}{\sin^{2}\theta}=-\frac{\alpha r^2}{8\pi\left(\alpha+r^{2}/l_{\rm P}^{2}\right)^{2}}\psi^{2}(1-C).
\end{equation}
Self-consistent solutions of the semiclassical field equations are sensitive to the regularity of the RSET acting as a source, even in situations where the classical spacetime prior to the backreaction has a physical singularity at $r=0$. In \cite{Arrechea2019} we obtained the semiclassical counterpart of the Schwarzschild geometry making use of the RP-RSET. We found that the counterpart to the Schwarzschild black hole has its horizon replaced by a wormhole neck. This neck is always placed above the Schwarzschild radius of the geometry. The RP-RSET allowed to extend the space of solutions to those whose neck lies below the Planck length, something forbidden for the Polyakov RSET due to its inherently singular form.

Before turning to the full semiclassical analysis, let us comment briefly on the regularity of the RP-RSET at horizons. To do so, let us calculate the semiclassical energy density \label{Eq:SemiDens} generated by a geometry with a Schwarzschild-like horizon. By Schwarzschild-like, we mean a local behaviour of the form 
\begin{equation}
   e^{2\phi} \propto \frac{r-r_{\rm H}}{r_{\rm H}}+\mathcal{O}\left(\frac{r-r_{\rm H}}{r_{\rm H}}\right)^{2}, \qquad
   1-C \propto \frac{r-r_{\rm H}}{r_{\rm H}}+\mathcal{O}\left(\frac{r-r_{\rm H}}{r_{\rm H}}\right)^{2},
\end{equation}
for the metric functions. Such profiles have a divergent RP-RSET with the energy density and pressure behaving as
\begin{equation}
   \rho_{\text{s}}=-p_{\text{s}}\propto-\frac{l_{\rm P}^{2}}{r_{\rm H}^{3}(r-r_{\rm H})}+\mathcal{O}\left(\frac{r-r_{\rm H}}{r_{\rm H}}\right)^{0}.
\label{eq:rho}
\end{equation}
It is expected that, once \eqref{Eq:Einstein} are solved at the self-consistent level, the backreaction of vacuum polarization on Schwarzschild-like horizons will be non-perturbative, thus modifying the metric so as these horizons disappear. This is a defining characteristic of the Boulware vacuum state, which is ill-defined at horizons (of the Schwarzschild kind). Had we considered the Hartle-Hawking vacuum instead, then the local structure around the horizon would have remained unspoiled, but at the cost of modifying the asymptotic regions.

Now consider a geometry that has an extremal horizon, characterized by the metric functions having a double zero at $r_{\rm H}$. An extremal horizon has zero surface gravity and, in consequence, the state-dependent terms of the Polyakov RSET \eqref{eq:RSETcomponents} vanish for all vacuum states, which become degenerate at the extremal horizon. We can easily check that the following profiles 
\begin{equation}\label{Eq:ExtrHor}
   e^{2\phi} \propto \left(\frac{r-r_{\rm H}}{r_{\rm H}}\right)^{2}+\mathcal{O}\left(\frac{r-r_{\rm H}}{r_{\rm H}}\right)^{3}, \qquad
   1-C \propto  \left(\frac{r-r_{\rm H}}{r_{\rm H}}\right)^{2}+\mathcal{O}\left(\frac{r-r_{\rm H}}{r_{\rm H}}\right)^{3},
\end{equation}
provide finite semiclassical density and pressures
\begin{equation}\label{eq:rho}
   \rho_{\text{s}}=p_{\text{s}} \propto\frac{l_{\rm P}^{2}}{r_{\rm H}^{4}}+\mathcal{O}\left(\frac{r-r_{\rm H}}{r_{\rm H}}\right),
\end{equation}
indicating that backreaction around an extremal configuration would, presumably, preserve the horizon. Indeed, later we will see that the horizon is preserved, although it receives non-perturbative quantum corrections that make it singular.
\section{Self-consistent electro-vacuum semiclassical equations}
\label{Sec:Self-consistent}

The following sections will be devoted to writing down the semiclassical Einstein equations \eqref{Eq:Einstein} having as sources the RP-RSET and the SET of the electromagnetic field, as well as discussing some of their most salient properties.  By analyzing these expressions we are able to determine the existence of three types of semiclassical solutions depending on the charge-to-mass ratio of the geometry (as in the Reissner-Nordström family), and reconstruct the shape of the solutions living in each of these three regimes. The $tt$ and $rr$ components of the semiclassical field equations are, respectively,
\begin{align}\label{Eq:ttsemi}
C+rC'
&
=\frac{Q^{2}}{r^{2}}+\frac{l_{\rm P}^{2}r^{2}}{r^{2}+\alpha l_{\rm P}^{2}}\left\{\left[2\psi'+\psi^{2}\right](1-C)-\psi C'\right\},\\
-2r\psi+C(1+2r\psi)\label{Eq:rrsemi}
&
=\frac{Q^{2}}{r^{2}}+\frac{l_{\rm P}^{2}r^{2}}{r^{2}+\alpha l_{\rm P}^{2}}\psi^{2}(1-C).
\end{align}
Instead of working with these two equations simultaneously, we can solve algebraically for $C$ in the second equation and plug the obtained expression into the first equation, which results into a first-order differential equation for the variable $\psi$:
\begin{align}\label{Eq:eqdif}
\psi'=
&
A_{0}+A_{1}\psi+A_{2}\psi^{2}+A_{3}\psi^{3},\quad \text{with}\nonumber\\
A_{0}(r)=
&
~Q^{2}\mathcal{D}(r),\nonumber\\
A_{1}(r)=
&
~2r\left[r^{2}-Q^{2}\left(2+\frac{l_{\rm P}^{2}}{2\left[r^{2}+\alpha l_{\rm P}^{2}\right]}\right)\right]\mathcal{D}(r),\nonumber\\
A_{2}(r)=
&
~r^{2}\left[2\left(r^{2}-Q^{2}\right)\left(1+\frac{l_{\rm P}^{2}r^{2}}{2\left[r^{2}+\alpha l_{\rm P}^{2}\right]^{2}}\right)+\frac{l_{\rm P}^{2}(2r^{2}-5Q^{2})}{r^{2}+\alpha l_{\rm P}^{2}}\right]\mathcal{D}(r),\nonumber\\
A_{3}(r)=
&
~\frac{r^{3}l_{\rm P}^{2}}{r^{2}+\alpha l_{\rm P}^{2}}\left[r^{2}\left(1+\frac{\alpha l_{\rm P}^{4}}{\left[r^{2}+\alpha l_{\rm P}^{2}\right]^{2}}\right)-Q^{2}\left(1+\frac{l_{\rm P}^{2}[r^{2}+2\alpha l_{\rm P}^{2}]}{\left[r^{2}+\alpha l_{\rm P}^{2}\right]}\right)\right]\mathcal{D}(r),\nonumber\\
\mathcal{D}(r)=
&\mbox{$\displaystyle-\frac{r^{2}+\alpha l_{\rm P}^{2}}{r^{2}\left(r^{2}-Q^{2}\right)\left[r^{2}+l_{\rm P}^{2}(\alpha-1)\right]}$}.
\end{align}
Once the solutions of this differential equation are analyzed, we can come back to the second equation in \eqref{Eq:ttsemi} in order to directly obtain the behavior of $C$.

In view of the above expression, it becomes clear that taking $\alpha>0$ is not sufficient for regularity, since $\mbox{$1/$}\mathcal{D}$ vanishes at $r=l_{\rm P}\sqrt{1-\alpha}$. By taking $\alpha>1$, we move this singularity outside of the domain of the radial coordinate, as it picks up an imaginary component. It is also interesting to note that \eqref{Eq:eqdif} is, in principle, ill-defined at $r=Q$ as well, as $\mbox{$1/$}\mathcal{D}$ vanishes for this radius too. 

The right-hand side of Eq. \eqref{Eq:eqdif} is a cubic polynomial in $\psi$ and can be factorized in roots that are functions of $r$. These roots are defined piecewise and can be matched along different intervals of the radial coordinate so that, when plotted, they appear as continuous curves. We have adopted the following definition
\begin{equation}\label{Eq:Roots}
\mathcal{R}_{1}=
\begin{cases} 
      \mathcal{S}_{1} & r\leq Q \\
      \mathcal{S}_{2} &  Q < r \leq r_{\text{i}}^{-} \\
      \mathcal{S}_{3} &  r_{\text{i}}^{-} < r \leq r_{\text{div}} \\
      \mathcal{S}_{1} & r> r_{\text{div}} 
   \end{cases}
,\quad
\mathcal{R}_{2}=
\begin{cases} 
      \mathcal{S}_{3} & r\leq r_{\text{i}}^{-} \\
	\mathcal{S}_{2} &  r> r_{\text{i}}^{-}
   \end{cases}
\nonumber\\
,\quad
\mathcal{R}_{3}=
\begin{cases} 
       \mathcal{S}_{2} & r\leq Q \\
       \mathcal{S}_{1} &  Q < r \leq r_{\text{div}} \\
      \mathcal{S}_{3} & r> r_{\text{div}} 
   \end{cases},
\end{equation}
where the expressions for $\mathcal{S}_{i}$ are, in terms of the coefficients in Eq. \eqref{Eq:eqdif},
\begin{align}
\mathcal{S}_{1}=
&
-\frac{A_{2}}{3A_{3}}\left[1+\frac{2^{1/3}\left(3A_{1}A_{3}-A_{2}^{2}\right)}{A_{2}\mathcal{H}}-\frac{\mathcal{H}}{2^{1/3}A_{2}}\right],
\nonumber\\
\mathcal{S}_{2,3}=
&
-\frac{3\pm i\sqrt{3}}{2}\left(1+\frac{A_{2}}{3A_{3}}\right)\mathcal{R}_{1},
\end{align}
with
\begin{align}
\mathcal{H}=
&
\left[-2A_{2}^{3}+9A_{1}A_{2}A_{3}-27A_{3}A_{0}\vphantom{\sqrt{\left(2A_{2}^{3}-9A_{1}A_{2}A_{3}+27A_{3}^{2}A_{0}\right)^{2}-4\left(A_{2}^{2}-3A_{1}A_{3}\right)^{3}}}\right.\nonumber\\
&
\left.+\sqrt{\left(2A_{2}^{3}-9A_{1}A_{2}A_{3}+27A_{3}^{2}A_{0}\right)^{2}-4\left(A_{2}^{2}-3A_{1}A_{3}\right)^{3}}\right]^{1/3}.
\end{align}
The symbols $r_{\text{i}}^{\pm}$ in \eqref{Eq:Roots} mark the lower and upper limits of a region where the roots acquire a non-zero complex part. Particularly, 
$\mathcal{R}_{1}$ and $\mathcal{R}_{2}$ become complex conjugate roots within this interval. The occurrence of complex roots has no impact on solutions, since, as we will prove in the following sections, solutions of \eqref{Eq:eqdif} with support in the interval $r\in(Q,r_{\text{div}})$ have to intersect certain fixed points.

The roots  $\{\mathcal{R}_i\}_{i=1}^3$ will be of utility in constraining the shape of solutions, since they indicate the turning points of the solution $\psi$. It is illustrative to compare with the analysis presented in \cite{Arrechea2019} of the semiclassical Schwarzschild counterpart, which is obtained taking $Q=0$ in the expressions above. For the Schwarzschild counterpart, one of the three roots is trivial $(\mathcal{R}=0)$, while the others always take negative values. The presence of charge introduces a non-zero zeroth-order term term in \eqref{Eq:eqdif}, $A_{0}\neq0$, modifying the shape of the roots (in particular, these now take positive values) and, in consequence, the domain of the solutions. 

\begin{figure}
\centering
\includegraphics[width=0.8\columnwidth]{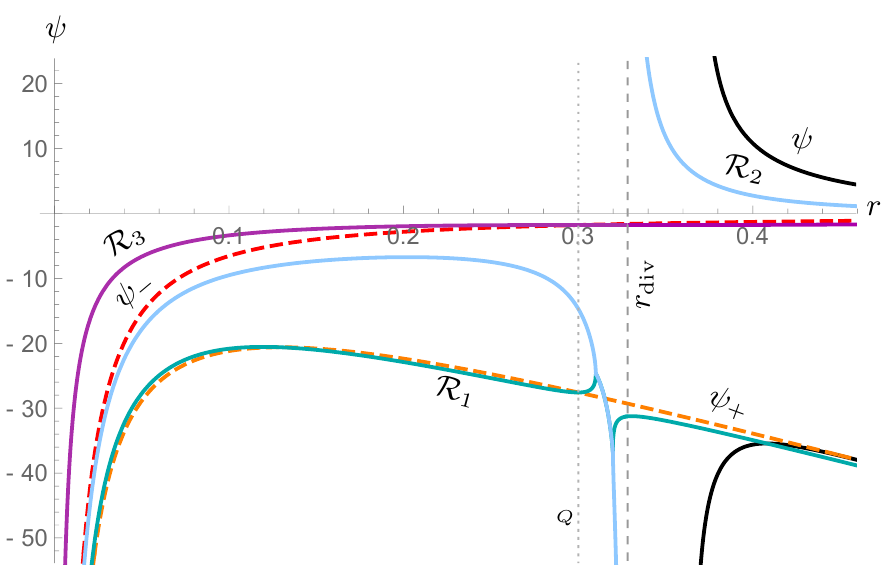}
\caption{Plot of the roots $\mathcal{R}_{1,2,3}$ (continuous curves) and the exact solutions $\psi_{\pm}$ (dashed lines). The roots are defined piecewise, and take negative values except for the positively diverging portion of $\mathcal{R}_{2}$. Its asymptote at $r=r_{\text{div}}$ (vertical, dashed line) marks the separatrix solution between the sub-extremal and super-extremal regimes. The dotted vertical line is $r=Q$. In this figure, we have taken $\alpha=1.01$ and $Q=0.3$ for visualization purposes. A numerical solution $\psi$ describing a wormhole of asymptotic mass $M=0.285$ has been drawn in black. The procedure giving rise to such solution is detailed in later sections.}
\label{Fig:Roots}
\end{figure}

In Fig. \ref{Fig:Roots} we show a plot of the roots and two exact solutions (the meaning of which is discussed right below) for a particular value of the charge parameter. The two special exact solutions of equation \eqref{Eq:eqdif} are
\begin{equation}\label{Eq:exactsol}
\psi_{\pm}=-\frac{r^{2}+\alpha l_{\rm P}^{2}}{l_{\rm P}^{2} r}\left(1\pm\sqrt{\frac{r^{2}+(\alpha-1)l_{\rm P}^{2}}{r^{2}+\alpha l_{\rm P}^{2}}}\right).
\end{equation}
The simplest way to find these exact solutions is analyzing the $C\rightarrow-\infty$ limit of the system of semiclassical equations; direct substitution shows that these are indeed actual exact solutions of \eqref{Eq:eqdif}. Remarkably, these are also exact solutions in the $Q=0$ situation that remain unchanged for nonzero values of the charge. In fact, we can rewrite \eqref{Eq:eqdif} as 
\begin{equation}\label{Eq:EqDifSimp}
\psi'=\mathcal{F}_{\text{Sch}}(r,\psi)+\mathcal{G}(r,Q,\psi)(\psi-\psi_{-})(\psi-\psi_{+}),
\end{equation}
where $\mathcal{F}_{\text{Sch}}$ corresponds to the right hand side of \eqref{Eq:eqdif} evaluated at $Q=0$, and
\begin{equation}\label{Eq:Gcoeff}
\mathcal{G}=\frac{l_{\rm P}^{2}Q^{2}\left(1+\frac{\displaystyle l_{\rm P}^{2}r}{\displaystyle r^{2}+\alpha l_{\rm P}^{2}}\psi\right)}{(r^{2}-Q^{2})\left[r^{2}+(\alpha-1)l_{\rm P}^{2}\right]}.
\end{equation}
This shows explicitly why $\psi_\pm$ can be exact solutions of \eqref{Eq:eqdif} for arbitrary values of the charge $Q$ even if they do not depend on $Q$, given that the $Q$-dependent terms in \eqref{Eq:eqdif} vanish identically for $\psi=\psi_\pm$. 

The presence of a double $\pm$ sign in \eqref{Eq:exactsol} comes from the existence of two branches of solutions of the semiclassical equations. These arise as a consequence of the semiclassical corrections introducing terms quadratic in $\psi$ in \eqref{Eq:rrsemi}. The branch associated with the $-$ sign (which we shall call ``unconcealed") returns the correct classical field equations in the $l_{\rm P}\to0$ limit, whereas the other (``concealed") is intrinsically non-perturbative and has no well-defined classical limit. Smooth transitions between both branches take place in certain situations, resulting in solutions that are more than a mere one-parameter deformation of the classical spacetime. There are two distinct situations that indicate the presence of a transition between branches: either the function $\psi$ jumps from $+\infty$ to $-\infty$, or the quantity $\mathcal{G}$ \eqref{Eq:Gcoeff} vanishes at some finite radius. These conditions follow from the requirement that the discriminant of Eq. \eqref{Eq:rrsemi} (seen as a second-order polynomial for $\psi$) vanishes.

We wish to end this section by stressing that, since \eqref{Eq:eqdif} is a first order differential equation for $\psi$, we can apply the uniqueness and existence theorem wherever $r\neq Q$; note that for $r\rightarrow Q$ the right-hand side of \eqref{Eq:EqDifSimp} diverges as $(r-Q)^{-1}$, which makes the differential equation singular there. This implies that solutions of this equation cannot intersect at finite $r$ but at the surface $r=Q$. In turn, we will show that the exact solutions \eqref{Eq:exactsol} act as boundaries for the remaining solutions. In addition, the roots \eqref{Eq:Roots} denote the turning points of $\psi$. These features will allow us to characterize the solutions of \eqref{Eq:eqdif} thoroughly.

\subsection{Asymptotically flat regime}
\label{Subsec:Asymptotic}

Let us start analyzing the behavior of \eqref{Eq:eqdif} at large radial distances and imposing asymptotic boundary conditions in order to select the solutions describing asymptotically flat spacetimes. The expectation value that leads to the RP-RSET is taken in the Boulware vacuum state, which has no particle content for asymptotic stationary observers. This ensures the existence of solutions in which semiclassical corrections decay sufficiently fast with distance as to preserve asymptotic flatness.

If we assume that the metric in \eqref{Eq:LineElement} is asymptotically flat, $\psi$ being the derivative of $\log(g_{tt})/2$ implies that the leading behavior at large distances must be given by
\begin{equation}\label{Eq:ansatz}
\psi\propto r^{-\eta},\qquad \eta\geq 2.
\end{equation}
Inserting this ansatz into \eqref{Eq:eqdif} and expanding in the large $r$ limit we obtain the following leading-order expansion
\begin{equation}\label{Eq:eta}
\psi'\propto-2r^{-\eta-1},
\end{equation}
which implies that the semiclassical equations are compatible with solutions with a leading behavior $\eta=2$, which are in fact associated with the leading terms in the asymptotic expansion of the Schwarzschild metric, the mass $M$ being the corresponding proportionality constant (up to numerical factors). 

Now, we allow the next subleading terms to enter expression \eqref{Eq:eta}, informing us about the first $Q$-dependent corrections,
\begin{equation}\label{Eq:RNeq}
\psi'\simeq-\frac{2\psi}{r}\left(1+r\psi\right)+\frac{Q^{2}}{r^{4}}.
\end{equation}
Integrating this expression twice yields
\begin{equation}\label{Eq:SolAsympt}
\phi\simeq\psi_{0}+\frac{1}{2}\ln\left[\text{cosh}\left(\frac{\sqrt{2}Q}{r}+\psi_{1}\right)\right],
\end{equation}
which now displays the leading asymptotic behavior characteristic of the Reissner-Nordström metric, in which
\begin{equation}
e^{2\phi}\simeq1-\frac{2M}{r}+\frac{Q^{2}}{r^{2}}+\mathcal{O}(1/r^3).
\end{equation}
In fact, in order to reproduce the above expression one just needs to choose the integration constants $\psi_0$ and $\psi_1$ as
\begin{equation}
\psi_{0}=\frac{1}{2}\ln\left[\text{sech}\left(\psi_{1}\right)\right],\quad \psi_{1}=-\text{arctanh}\left(\frac{\sqrt{2}M}{Q}\right),
\end{equation}
Note that the expression of the leading-order contribution \eqref{Eq:ansatz} serves to determine which terms in Eq. \eqref{Eq:eqdif} enter the expansion \eqref{Eq:RNeq} at the same order as the first $Q$-dependent term. Assuming that $\psi$ obeys an expansion with additional terms decaying with $r$, as in Eq. \eqref{Eq:ansatz}, is not sufficient to find the correct asymptotic limit. Such an expansion for $\psi$ fails to properly approximate the asymptotic solution since, for the Reissner-Nordstöm geometry, $\psi$ equals a quotient between polynomials. At best, expanding $\psi$ in powers of $r$ succeeds in returning the Schwarzschild geometry, but fails in determining the Reissner-Nordström geometry, which involves subleading contributions. As a consequence, the correct asymptotic metric comes from solving the approximate expression \eqref{Eq:RNeq} in a self-consistent manner.

In addition, the behaviour for $C$ is found by replacing \eqref{Eq:SolAsympt} inside \eqref{Eq:rrsemi} and taking the leading order contribution for large $r$. We obtain $(1-C)\simeq e^{-2\phi}$ in the asymptotic limit, so the Reissner-Nordström geometry is fully recovered.

After fixing the asymptotic behavior of the different solutions, we proceed by integrating the semiclassical equations inwards, towards smaller values of $r$. In doing so, several scenarios can arise depending on the balance between the charge $Q$ and the asymptotic mass $M$. We start our inwards integration from the asymptotic region with a positive $\psi$ (this condition is equivalent to assuming $M>0$ since, by virtue of \eqref{Eq:rrsemi}, positivity of $\psi$ ensures positivity of $C$) that situates the solution above all roots and analytical exact solutions depicted in Fig. \ref{Fig:Roots}. Self-consistency of \eqref{Eq:eqdif} ensures that $\psi$ grows monotonically inwards unless it intersects one of the roots. In view of Fig. \ref{Fig:Roots}, there exist three posibilities: the solution $\psi$ either diverges at some radius $r>r_{\text{div}}$; it grows sufficiently slow as to cross $r_{\text{div}}$, encountering a maximum and extending to $r\simeq0$; or it stays in between both regimes, diverging at $r=r_{\text{div}}$ at the same rate as the root $\mathcal{R}_{2}$ does. 
The solution will follow either of these paths depending on the relative values of $Q, M$ and $\alpha$. In turn, we can assure that, as long as $M>0$ and $\alpha>1$, there exists a critical value of the charge $Q_{\text{crit}}$ that corresponds to the separatrix solution. In the following we analyze these three cases individually.

\section{Wormhole geometry}
\label{Sec:Wormhole}

The first solutions we analyze are deformed continuously to the Schwarzschild case in the limit $Q\rightarrow0$, and are valid up to a critical value of the charge $Q_{\rm crit}$ that is determined by the self-consistency of these solutions.

\subsection{Near-neck expansion}

As the first step in our analysis, we assume a positive value of $\psi$ at a large initial radius. This starting assumption will be present in all the following sections, whereas the particularities of the $\psi<0$ case will be detailed at due time. By self-consistency of \eqref{Eq:eqdif}, $\psi$ grows monotonically inwards until it either reaches the root $\mathcal{R}_{2}$, diverges exactly at $r=r_{\text{div}}$, or it does so at some radius $r>r_{\text{div}}>Q$. This last possibility is the one we explore in the present section. Assuming that the function $\psi$ diverges at some finite radius $r_{\rm{B}}>r_{\text{div}}$ (the value $r_{\rm{B}}$ stands for bouncing surface of the radial function, as we will see below), the differential equation \eqref{Eq:eqdif} can be approximated, at leading order in $\psi$, by
\begin{equation}\label{Eq:A3}
\psi'\simeq A_{3\rm B}\psi^{3}.
\end{equation}
The term $A_{3\rm B}$ is just the coefficient $A_{3}$ in \eqref{Eq:eqdif} evaluated in the $r\to r_{\rm B}$ limit, where it takes a constant value
\begin{equation}
A_{3\rm B}=-\frac{l_{\rm P}^{2}r_{\rm B}\left[r_{\rm B}^{2}\left(1+\frac{\displaystyle\alpha l_{\rm P}^{4}}{\displaystyle\left[r_{\rm B}^{2}+\alpha l_{\rm P}^{2}\right]^{2}}\right)-Q^{2}\left(1+\frac{\displaystyle l_{\rm P}^{2}\left[r_{\rm B}^{2}+2\alpha l_{\rm P}^{2}\right]}{\displaystyle\left[r_{\rm B}^{2}+\alpha l_{\rm P}^{2}\right]^{2}}\right)\right]}{\left(r_{\rm B}^{2}-Q^{2}\right)\left[r_{\rm B}^{2}+l_{\rm P}^{2}(\alpha-1)\right]}.
\end{equation}
The sign of this constant depends on the value of the charge $Q$. Integrating the differential equation \eqref{Eq:A3} returns the following pair of solutions
\begin{equation}\label{Eq:PsiNeckSimp}
\psi\simeq\pm\sqrt{\frac{k_{0}}{4(r-r_{\rm B})}}+\mathcal{O}\left(r-r_{\rm B}\right)^{1/2},
\end{equation}
where, for consistency with the notation in \cite{Arrechea2019}, we have introduced the redefinition
\begin{equation}
k_{0}=-\frac{2}{A_{3\rm B}}.
\end{equation}
Here, the integration constant has been fixed to preserve the divergence in $\psi$ as $r\to r_{\rm {B}}$. The $\pm$ signs indicate that there are two solutions, one per branch. For the moment, we take the $+$ sign since, in the inwards integration, $\psi$ diverges towards positive infinity. Later, we will give meaning to the $-$ sign solution as well.

For each of these two solutions to be real, the constant $k_0$ must be positive, which in turn implies that
\begin{equation}\label{Eq:under_q}
\abs{Q}<r_{\rm B}\sqrt{\frac{\alpha l_{\rm P}^{4}+\left(r_{\rm B}^{2}+\alpha l_{\rm P}^{2}\right)^{2}}{l_{\rm P}^{2}\left(r_{\rm B}^{2}+2\alpha l_{\rm P}^{2}\right)+\left(r_{\rm B}^{2}+\alpha l_{\rm P}^{2}\right)^{2}}}.
\end{equation}
This is the semiclassical equivalent of the condition that guarantees the non-extremal nature of the classical Reissner-Nordström black hole. We will analyze in more detail the content of this constraint on the charge below; for the moment let us keep describing the elements of the metric.

The redshift function follows from integrating Eq. \eqref{Eq:PsiNeckSimp} 
\begin{equation}\label{Eq:RedsNeck}
\phi\simeq\sqrt{k_{0}(r-r_{\rm B})}+\phi_{\rm B}+\mathcal{O}\left(r-r_{\rm B}\right)^{3/2},
\end{equation}
where the integration constant $\phi_{\rm B}$ denotes the value of the redshift function at $r=r_{\rm B}$. This parameter must be found by numerically integrating the solution from the asymptotic region until the surface $r_{\rm B}$ is encountered.

As for the compactness function, replacing Eq. \eqref{Eq:PsiNeckSimp} inside \eqref{Eq:rrsemi} returns the compactness in the $r\to r_{\rm B}$ limit, which goes as
\begin{equation}\label{Eq:CompNeck}
C\simeq1-k_{\rm 1} (r-r_{\rm B})+\mathcal{O}\left(r-r_{\rm B}\right)^{3/2},
\end{equation}
with 
\begin{equation}
k_{1}=\frac{4(r_{\rm B}^{2}-Q^{2})(r_{\rm B}^{2}+\alpha l_{\rm P}^{2})}{l_{\rm P}^{2}r_{\rm B}^{2}k_{0}}>0.
\end{equation}
Observing the metric functions \eqref{Eq:RedsNeck} and \eqref{Eq:CompNeck} it is noticeable that, at $r=r_{\rm B}$, the redshift function is finite and nonzero, while compactness equals $1$. As a consequence, the radial component of the metric diverges at the spherical surface $r_{\rm B}$, while the time component remains finite, indicating the pressence of a bouncing surface for the radial coordinate. Schwarzschild coordinates do not provide an adequate description of such surfaces, so we perform a coordinate transformation to the proper coordinate $l$
\begin{equation}\label{Eq:PropCoord}
\frac{dl}{dr}=\pm\frac{1}{\sqrt{k_{\rm 1}(r-r_{\rm B})}},
\end{equation}
where the $\pm$ sign informs us about the branch of the radial coordinate in which observers are located. Integrating \eqref{Eq:PropCoord} yields
\begin{equation}\label{Eq:PropCoordInt}
l-l_{\rm B}=\pm\sqrt{\frac{4(r-r_{\rm B})}{k_{1}}}.
\end{equation}
Therefore,  the radial coordinate has a minimum at the surface $r_{\rm B}$, while the $l$ coordinate is continuous at that minimal surface. The $+$ sign
in \eqref{Eq:PropCoord} denotes the side of this minimal surface where $r$ grows with $l$, that is, the side that contains the asymptotically flat region. The $-$ sign denotes the other side of $r_{\rm B}$, where $r$ decreases as $l$ grows. A jump between the two branches of solutions in \eqref{Eq:PsiNeckSimp} takes place at $l=l_{\rm B}$. 
In terms of the $l$ coordinate, the line element around $r_{\rm B}$ is given by
\begin{equation}\label{Eq:NeckMetric}
ds^{2}\simeq-e^{\sqrt{k_{0}k_{1}}(l-l_{\rm B})+2\phi_{\rm B}}dt^{2}+dl^{2}+\left[\frac{k_{1}}{4}\left(l-l_{\rm B}\right)^{2}+r_{\rm B}^{2}\right]^{2}d\Omega^{2}.
\end{equation}
In the above expression, we have integrated \eqref{Eq:PsiNeckSimp} with the $\pm$ sign and performed the coordinate change \eqref{Eq:PropCoordInt}, matching the signs so that the line element is continuous and differentiable through $l=l_{\rm B}$.
The line element \eqref{Eq:NeckMetric} represents the metric in the near-neck region of an asymmetric wormhole. The $l>l_{\rm B}$ region connects smoothly with the Reissner-Nordström geometry sufficiently far from the neck, whereas the $l<l_{\rm B}$ domain corresponds to a new asymptotic region with a radically different asymptotic structure. 
As we already showed in \cite{Arrechea2019}, the semiclassical equations in the Boulware vacuum do not permit the existence of non-extremal horizons, meaning horizons with a non-zero surface gravity. Here we show that the charged case exhibits the same type of elimination of the horizon: substitution by a wormhole neck. The cubic dependence in $\psi$ introduced by the RP-RSET in \eqref{Eq:eqdif} is the ingredient that ultimately causes the occurrence of these horizonless solutions. In physical terms, this is a consequence of the huge backreaction that takes place when non-extremal horizons and quantum fields living in the Boulware vacuum state are forced to coexist. Understood as an integration from the asymptotic region, the mass contribution provided by the vacuum polarization of the scalar field results in the compactness reaching $1$ at a faster rate than vanishing of the redshift function. This contribution amounts to a cloud of negative mass that extends to infinity. The role of the charge here is, similarly to the classical situation, to produce a repulsive contribution that slows down the inwards growth rate of the Misner-Sharp mass. Thus, the neck radius is dragged towards smaller values of $r$ as $Q$ increases. 

\begin{figure}
\centering
\includegraphics[width=0.8\columnwidth]{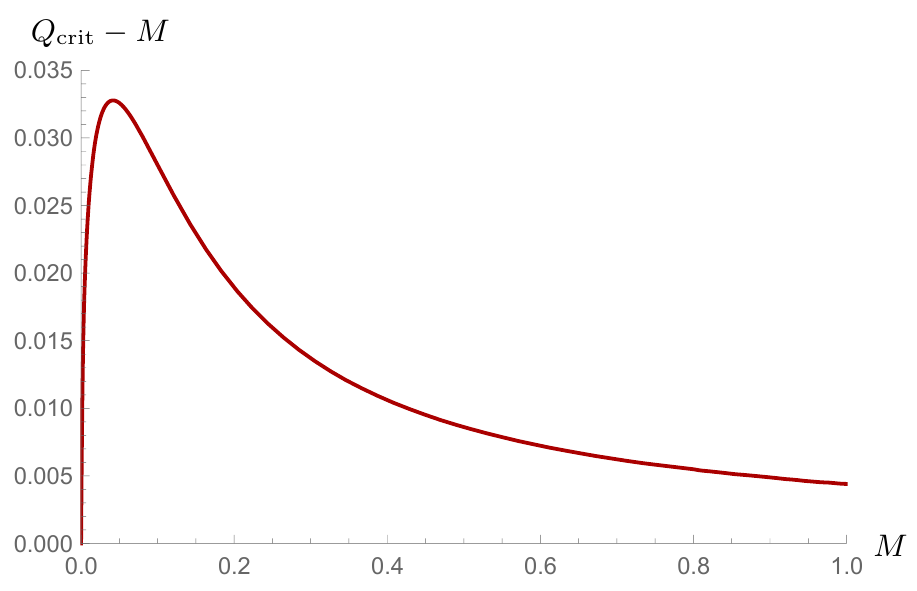}
\caption{Plot of the quantity $Q_{\text{crit}}-M$ in terms of $M$. The parameter $Q_{\text{crit}}$ denotes the value of the charge that separates under-charged from over-charged spacetimes.}
\label{Fig:QCrit}
\end{figure}

In the $Q\to0$ limit, this geometry connects smoothly with the semiclassical Schwarzschild counterpart, which retains the same wormhole structure. On the other hand, the solution \eqref{Eq:PsiNeckSimp} provides an upper bound for the charge parameter given by \eqref{Eq:under_q}. Since the parameter $r_{\rm B}$ can only be found numerically, we are unable to provide an exact analytical bound for $Q$ in terms of the asymptotic mass $M$ without resorting to numerical calculations. In Figure \ref{Fig:QCrit} we have plotted the first value of $Q$ that takes the solution out of the wormhole regime for several (although quite small in Planck units) asymptotic masses. 
We find that the separatrix charge $Q_{\text{crit}}$ needs to surpass the asymptotic mass $M$. This is so because the vacuum polarization of the scalar field, by backreacting on the geometry, modifies the mass of the spacetime, while charge stays constant throughout the geometry. In order to compensate for this increase in mass as smaller radii are approached, $Q_{\text{crit}}$ lies above its classical value $Q_{\text{crit}}=M$, as the expansion of \eqref{Eq:under_q} around its classical value shows
\begin{equation}
\abs{Q}\lesssim r_{\rm B}\left[1-\frac{l_{\rm P}^{2}}{2 r_{\rm B}^{2}}+\mathcal{O}\left(\frac{l_{\rm P}}{r_{\rm B}}\right)^{4}\right],
\end{equation}
where $r_{\rm B}>r_{+}$ always. Since for classical sub-extremal black holes $r_{+}>M$, we conclude that the smallest value of $Q$ that does not obey \eqref{Eq:under_q} must be greater than $M$.

We want to end this section with some remarks on the validity of our approximate solutions. Two situations can be distinguished: for big wormhole throats $(r_{\rm B}\gg\sqrt{\alpha}l_{\rm P})$, the metric \eqref{Eq:NeckMetric} remains valid as long as
\begin{equation}
r-r_{\rm B}\ll \frac{l_{\rm P}^{2}}{r_{\rm B}},
\end{equation}
whereas for small wormholes $(Q\ll r_{\rm B}\ll\sqrt{\alpha}l_{\rm P})$,
\begin{equation}
r-r_{\rm B}\ll r_{\rm B}
\end{equation}
must hold. At distances where the above expressions fail to be good approximations, the metric should be matched with the Reissner-Nordström line element or with the asymptotic metric at the inner side of the wormhole.

\subsection{Asymptotic singularity}

The metric on the other side of the neck can be uniquely determined solely by the roots and exact solutions $\psi_{\pm}$ depicted in Figure \ref{Fig:Roots}. Relying in the fact that just beyond the neck the solution is well-described by the $-$ sign in \eqref{Eq:PsiNeckSimp}, and that, necessarily, $r_{\rm B}>r_{\text{div}}$, it follows that $\psi$ takes values below all the roots and exact solutions. By self-consistency of \eqref{Eq:eqdif}, $\psi$ grows with $r$ until it encounters the root $\mathcal{R}_{1}$, which decreases almost linearly with $r$ (check the bottom right portion of Figure \ref{Fig:Roots}). As the root is crossed, $\psi$ begins decreasing monotonically confined between $\mathcal{R}_{1}$, which cannot be encountered for a second time, and the exact solution $\psi_{+}$, which cannot be crossed either from $r_{\rm B}$ to $\infty$, by virtue of the Picard-Lindelöf theorem.
 
The particular form of $\psi$ deep inside the neck can be determined analytically assuming that, asymptotically, the $\psi$ function goes as
\begin{equation}\label{Eq:psiasympt}
\psi=\psi_{+}+\beta(r),
\end{equation}
where $\beta(r)$ is a function measuring the deviation from the exact solution. Replacing inside \eqref{Eq:eqdif}, keeping terms up to linear order in $\beta$, and taking the limit $r\to\infty$ results in
\begin{equation}\label{Eq:BetaAsympt}
\beta'\simeq-\frac{\left\{-16r^{4}+8 l_{\rm P}^{2}\left[Q^{2}+r^{2}(1-2\alpha)\right]+l_{\rm P}^{4}(5-32\alpha)\right\}}{4 l_{\rm P}^{2} r^{3}}\beta,
\end{equation}
which integrated \eqref{Eq:BetaAsympt} gives
\begin{equation}
\beta= \beta_{0}\left(\frac{r}{l_{\rm P}}\right)^{1-4\alpha}e^{-\frac{2r^{2}}{l_{\rm P}^{2}}}\left[1-\frac{Q^{2}}{r^{2}}-\frac{ l_{\rm P}^{2}(5-32\alpha)}{8 r^{2}}+\mathcal{O}(r^{-4})\right],
\end{equation}
where $\beta_{0}$ is an integration constant with dimensions of inverse of length. Inserting this inside \eqref{Eq:psiasympt} and integrating to obtain the redshift function and compactness yields  the asymptotic metric
\begin{align}\label{eq:metricnopert}
    ds^{2}
    &
    \simeq
    \left(\frac{r}{l_{\rm P}}\right)^{1-4\alpha}e^{- {2r^{2}}/{l_{\rm P}^{2}}}\left\{-a_{0}\left(1-\frac{l_{\rm P}^{2}}{8 r^{2}}\right)dt^{2}
    +\frac{2\beta_{0} r^{2}}{l_{\rm P}}\left[1-\frac{(9-32\alpha)l_{\rm P}^{2}}{8r^{2}}\right]dr^{2}\right\}+r^{2}d\Omega^{2}.
\end{align}
$Q$-dependent terms enter this expression as subdominant contributions, so they do not alter the asymptotic form of the geometry, which is the same as in the Schwarzschild case. This becomes evident already in Eq. \eqref{Eq:eqdif}, where terms proportional to $Q$ are subdominant with respect to the ``vacuum", $r$-dependent contributions from the Schwarzschild sector. 
 
\begin{figure}
\centering
\includegraphics[width=0.75\columnwidth]{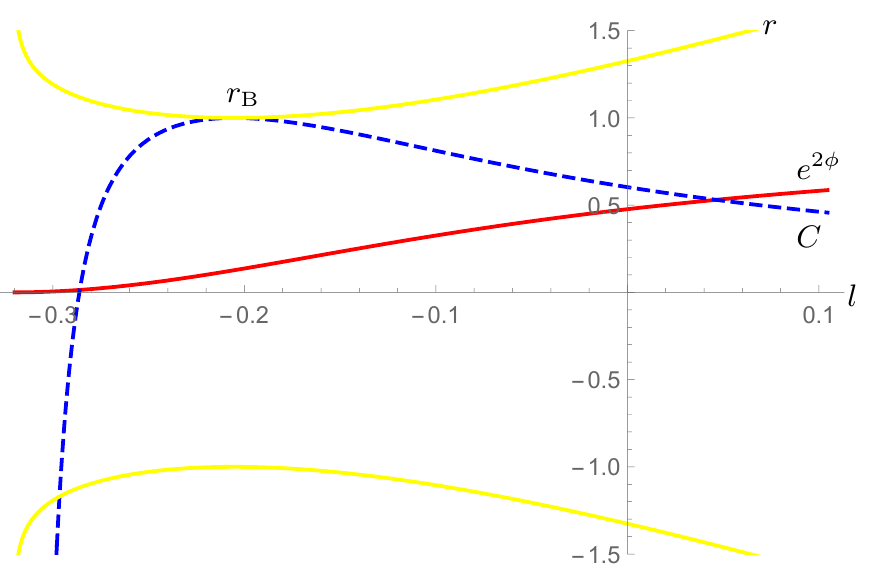}
\caption{Numerical plot of the wormhole geometry around the neck region. The yellow curves represent the radial function $r(l)$ (in units of $r_{\rm B}$). The red curve is the redshift function $e^{2\phi}$ and the blue, dashed line represents the compactness function $C$. The right hand portion of the diagram connects with the asymptotically flat region, where $e^{2\phi}=1-C=1$. The other side of the wormhole ends in a null singularity at $r\to\infty$. The proper distance between this null singularity and the neck is finite. The parameters from this simulation have been assigned small values equal to  $M=0.08, Q=0.05$ and $\alpha=1.01$ for visualization purposes, but the overall features of the solution remain unchanged for higher masses and charges.}
\label{Fig:WormholeNeck}
\end{figure}

Concerning the structure of the singularity, its main property is that it is a consequence of the runaway of vacuum polarization, which makes the Misner-Sharp mass diverge towards negative infinity. By inspection it is possible to see that the $r\to\infty$ limit is a null singularity. Moreover, this singularity is located at finite affine distance from the neck for all geodesic observers. Figure \ref{Fig:WormholeNeck} contains a numerical plot of the metric functions in terms of the proper coordinate $l$ for an internal region around the neck (the numerical integration goes beyond the analytical approximation). Then, the geometry connects an asymptotically flat region on the right-hand side with an asymptotic null singularity on the other end. From this perspective, there is no qualitative difference with respect to the Schwarzschild case ($Q=0$) discussed in \cite{Arrechea2019}. To wrap up this discussion, let us stress that whenever $M>Q_{\rm crit} \gg l_{\rm P}$ the results found would have been qualitatively the same if we would have used instead just the Polyakov approximation. In fact, because of the appearance of a bounce at $r_{\rm B} \gg l_{\rm P}$, the region close to $r=0$ of the approximation is not explored by most of the solutions found. As the Polyakov RSET is a good approximation to the RSET around horizons, we can expect that the presence of asymmetric wormhole solutions is a robust characteristic of the semiclassical electrovacuum equations. When $r_{\rm B} \sim l_{\rm P}$ we are entering into a much more slippery terrain. We cannot be sure whether the wormhole shape would be maintained in more refined approximations to the RSET. Notice, however, that this regime corresponds also to $M \sim l_{\rm P}$, and that it is reasonable to expect that the structure of black holes with Planckian size would surely lie outside the regime of validity semiclassical gravity.

In the following sections, we will turn to other non-wormhole geometries with values of the charge greater than or equal to the separatrix value $Q_{\text{crit}}$.

\section{Over-charged regime}
\label{Sec:Naked}
\begin{figure}
\centering
\includegraphics[width=0.8\columnwidth]{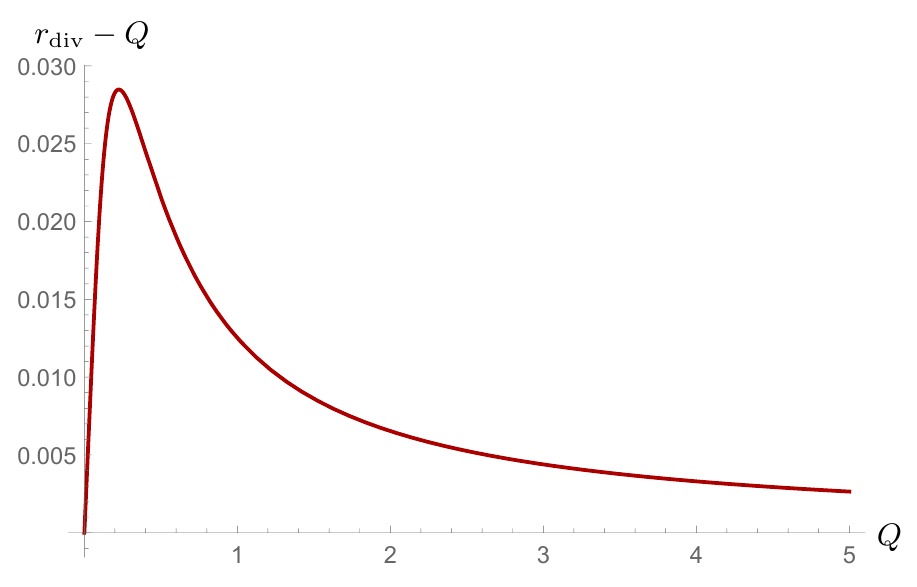}
\caption{Plot of the quantity $r_{\text{div}}-Q$ for various values of the charge $Q$. Whereas for large charges this quantity tends to zero, in the regime of small charges comparable, in magnitude, to $l_{\rm P}$, this difference increases appreciably, going to $0$ again in the $Q\to0$ limit.}
\label{Fig:rDiv}
\end{figure}

We now turn to the case where solutions have no wormhole neck and thus extend all the way down to $r=0$. There are two ways of accomplishing this behaviour: either considering a negative mass ($M<0$), or by increasing $Q$ beyond its critical value $Q_{\text{crit}}$ while $M>0$. In the latter case, compactness increases inwards, although its growth rate slows down as $Q$ increases. Similarly, the repulsion exerted by the electromagnetic field slows down the growth rate of $\psi$, displacing the surface $r_{\rm B}$ towards smaller radii. As $Q$ increases, the root $\mathcal{R}_{2}$, which diverges (while remaining positive) at $r_{\text{div}}$, is eventually intersected by the solution. Indeed, the explicit expression for $r_{\text{div}}$ can be derived from solving the polynomial expression $A_{3}=0$ for $r$. From all the roots that factorize this expression, the one which corresponds to the surface of infinite $\mathcal{R}_{2}$ turns out to be
\begin{align}\label{Eq:rdiv}
r_{\text{div}}=
&
\left\{2^{4/3}Q^{4}+2Q^{2}\left[Q^{2}\mathcal{V}+l_{\rm P}^{2}\left(3^{3/2}\mathcal{U}+\alpha \mathcal{V}\right)\right]^{1/3}+\left[2Q^{2}\mathcal{V}+2l_{\rm P}^{2}\left(3^{3/2}\mathcal{U}+\alpha \mathcal{V}\right)\right]^{2/3}\right.\nonumber\\
&
\left.+2l_{\rm P}^{2}\left[2^{1/3}Q^{2}(3+2\alpha)-2\alpha\left(Q^{2}\mathcal{V}+l_{\rm P}^{2}\left[3^{3/2}\mathcal{U}+\alpha\mathcal{V}\right]\right)^{1/3}\right]+2^{4/3}l_{\rm P}^{4}\alpha(\alpha-3)\right\}\nonumber\\
&
\times \left[6Q^{2}\mathcal{V}+6l_{\rm P}^{2}\left(3^{3/2}\mathcal{U}+\alpha\mathcal{V}\right)\right]^{1/6},
\end{align}
with
\begin{align}
\mathcal{U}=
&
\left\{Q^{8}(4\alpha-1)+2l_{\rm P}^{2}Q^{6}\left[\alpha(9+8\alpha)-2\right]+\alpha l_{\rm P}^{4}Q^{4}\left[12+\alpha(47+24\alpha)\right]\right.\nonumber\\
&
\left.+4\alpha^{2}l_{\rm P}^{6}Q^{2}\left[\alpha(9+8\alpha)-3\right]+4\alpha^{3}l_{\rm P}^{8}(1+\alpha)^{2}\right\}^{1/2},\nonumber\\
\mathcal{V}=
&
\ 2Q^{4}+l_{\rm P}^{2}Q^{2}(9+4\alpha)+2\alpha l_{\rm P}^{4}(9+\alpha).
\end{align}
In the $l_{\rm P}\to0$ limit, this quantity goes as 
\begin{equation}\label{Eq:rdivclas}
r_{\text{div}}\simeq Q+\frac{l_{\rm P}^{2}}{2Q}+\mathcal{O}\left(\frac{l_{\rm P}^{4}}{Q^{2}}\right),
\end{equation}
which imples that, in the classical limit, this surface lies at $r=Q$. Similar arguments as those presented here are valid in the analysis of the classical equations and, in that case, the surface $r=Q$ serves as a separatrix between sub-extremal Reissner-Nordström geometries, in which $\psi$ diverges at $r=r_{+}>Q$, and super-extremal Reissner-Nordström geometries, where $\psi$ is bounded from above. The solution where $\psi$ diverges exactly as $(r-Q)^{-1}$ corresponds to the extremal black hole. The introduction of the length scale $l_{\rm P}$ related to quantum corrections displaces the separatrix surface outwards. Figure \ref{Fig:rDiv} shows a plot of the quantity $r_{\text{div}}-Q$ in terms of the charge $Q$.

\begin{figure}
\centering
\includegraphics[width=0.8\columnwidth]{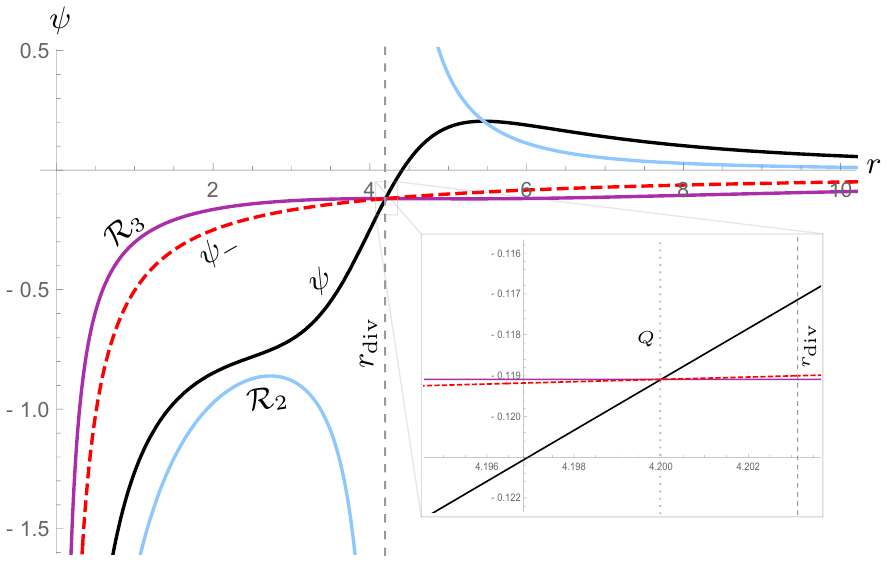}
\caption{Numerical plot of an over-charged solution (black curve) alongside the roots $\mathcal{R}_{2,3}$ and the exact solution $\psi_{-}$. Once it intersects $\mathcal{R}_{2}$, the solution must cross $\psi_{-}$ at $r=Q$ and becomes trapped between $\mathcal{R}_{2}$ and $\psi_{-}$ by self-consistency of the differential equation. If the maximum for $\psi$ is reached closer to $r_{\text{div}}$, then the solution decreases more abruptly and can intersect $\mathcal{R}_{2}$ zero, one or two times, but the behaviour at $r=0$ remains unchanged. For this plot we have chosen $M=4, Q=1.05M$ and $\alpha=1.01$. The region around the intersection point $r=Q$ has been depicted in detail. }
\label{Fig:NakedSing}
\end{figure}

The crossing with $\mathcal{R}_2$ at $r=r_{\rm div}$ marks a maximum in $\psi$ that prevents the geometry from acquiring a wormhole shape. Instead, the coordinate $r$ now extends to $r=0$. In doing so, $\psi$ has to cross $r=Q$, the surface where the differential equation \eqref{Eq:eqdif} is singular (check Fig. \ref{Fig:NakedSing}).
We can prove that the solution is regular there taking into account Eq. \eqref{Eq:EqDifSimp}. Both $\psi_{\pm}$ \eqref{Eq:exactsol} are compatible with a regular behaviour of the differential equation at $r=Q$, as for the $\psi_{\pm}$ solutions, the $Q$-dependent terms in \eqref{Eq:eqdif} (which contain the possible singularities) vanish identically. Any other solution must intersect $\psi_{+}$ or $\psi_{-}$ at $r=Q$ to avoid a singularity. Since the right-hand side of \eqref{Eq:EqDifSimp} is divergent, the Picard-Lindelöf theorem does not hold at $r=Q$, allowing solutions to intersect at that precise surface. Due to continuity, the solution we are describing in this section will intersect first $\psi_-$. Now let us derive the form of the solution $\psi$ around $r=Q$ by assuming
\begin{equation}\label{Eq:PsiGamma}
\psi=\psi_{-}+\xi(r),
\end{equation}
where $\xi$ is the function that measures the deviation between solutions. To guarantee the finiteness of \eqref{Eq:Gcoeff}, $\xi$ must vanish at least linearly as $r$ approaches $Q$. Replacing \eqref{Eq:PsiGamma} and its derivative in \eqref{Eq:EqDifSimp} and dropping terms beyond linear order in $\xi$, we obtain in the $r\to Q$ limit
\begin{equation}
\xi'(r)\simeq\frac{\xi(r)}{r-Q},
\end{equation}
which upon integration returns
\begin{equation}
\xi\simeq\left(r-Q\right)\xi_{0},
\end{equation}
where $\gamma_{0}$ is an integration constant with dimensions of inverse of length squared. Thus, as $\psi$ crosses $r=Q$, it does so in a manner that ensures that the right-hand side of \eqref{Eq:EqDifSimp} does not diverge. 

After the solution crosses $r=Q$ towards smaller radii, it can intersect $\mathcal{R}_{3}$ zero times, once or twice. In the first case, $\psi$ stays confined between the exact solution $\psi_{-}$ and the root $\mathcal{R}_{3}$, its value decreasing until it reaches $-\infty$, as depicted in Fig. \ref{Fig:NakedSing}. The situation in which there are two intersections with $\mathcal{R}_3$ occurs whenever $\psi$ decreases sufficiently fast (after crossing $r=Q$) so that it intersects the rightmost part of $\mathcal{R}_{3}$. After the first intersection, it grows until $\mathcal{R}_{3}$ is encountered for a second time. In between these two situations there is one in which $\mathcal{R}_{3}$ is touched once tangentially at the same point in which $\mathcal{R}_{3}$ reaches its maximum value, which becomes a saddle point of $\psi$. This qualitative description of the behavior of the solutions is therefore exhaustive.

The other posibility that we mentioned at the beginning of this section requires a negative asymptotic mass, $M<0$. In this case, $\psi$ takes negative values above $\psi_{-}$ at infinity. By self-consistency of \eqref{Eq:eqdif}, the solution decreases monotonically inwards, goes across $r_{\text{div}}$, and diverges towards $-\infty$ in the $r\to0$ limit, confined between $\mathcal{R}_{3}$ and $\psi_{-}$. Thus, the $r\simeq0$ behaviour of the over-charged solution and the geometry with negative asymptotic mass is the same. One of the shared features is the existence of a curvature singularity at $r=0$, as it is analyzed in the next section.

\subsection{Singularity at $r=0$}
\label{Subsec:sing}

The form of the solution close to the origin in the over-charged regime comes from a combination between the (singular) electromagnetic field and the effects of vacuum polarization. The magnitude of the RP-RSET describing quantum fluctuations near $r=0$ is strongly affected by the value of the regulator. Indeed, if the RP-RSET is sufficiently suppressed by considering a large value of $\alpha$, the dominant source of compactness in the $r\to0$ limit will come from terms proportional to the charge $Q$. In order to obtain the behaviour of the compactness function close to the radial origin we perform an asymptotic analysis of \eqref{Eq:ttsemi} in the $r\to0$ limit.

Firstly, in order to simplify \eqref{Eq:ttsemi}, we need to obtain the form of solutions of Eq. \eqref{Eq:eqdif} near the origin. The resulting approximate expression and its first derivative $\psi'$ are then inserted in \eqref{Eq:ttsemi}. By assuming $\psi$ takes the form 
\begin{equation}\label{Eq:PsiA}
\psi=\frac{a}{r},
\end{equation}
which is the only profile both divergent at $r\to0$ and compatible with Eq. \eqref{Eq:eqdif}. Replacing \eqref{Eq:PsiA} and its first derivative in Eq. \eqref{Eq:eqdif} and expanding in the $r\to0$ limit, we obtain the following solutions for the constant $a$:
\begin{equation}\label{Eq:as}
a_{0}=-\frac{\alpha}{2+\alpha},\quad a_{\pm}=-\left(\alpha\pm\sqrt{\alpha(\alpha-1)}\right).
\end{equation}
 The first of these values is a solution that appears as a consequence of the introduction of the electric charge. We have not been able to relate it with any of the situations described in this work, so chances are that this solution does not connect with an asymptotically flat region. The last two values correspond to evaluating the exact solutions $\psi_{\pm}$ in the $r\to0$ limit. The complete solution depicted in Fig. \ref{Fig:NakedSing} stays within the branch in which initial conditions have been imposed (in other words, imposing the condition of asymptotic flatness implies that the concealed branch is never explored), so we stay with the coefficient $a_{-}$ in \eqref{Eq:as}, which has a well-defined classical limit. Terms subdominant with respect to the leading order \eqref{Eq:PsiA} (which would be linear in $r$) can be obtained, but the leading-order form of $\psi$ is sufficient to illustrate our point in this section.

After substitution of \eqref{Eq:PsiA} and its derivative in \eqref{Eq:ttsemi}, we obtain the following approximate linear differential equation for the compactness
\begin{equation}\label{Eq:Compasympt}
 	C'=\frac{\left[(2-a_{-})a_{-}-\alpha\right]C}{\left(\alpha+a_{-}\right)r}\left[1+\mathcal{O}(r)\right]+\frac{\alpha Q^{2}}{(\alpha+a_{-})r^{3}}\left[1+\mathcal{O}(r^{-1})\right].
\end{equation}
In this equation, subleading terms divergent in $r$ that accompany the compactness and the charge can be neglected, since they contribute to the solution at subdominant order. Integrating Eq. \eqref{Eq:Compasympt} is straightforward, yielding
\begin{equation}\label{Eq:CompExp}
C=c_{1}\left(\frac{r}{l_{\rm P}}\right)^{b_{1}}\left[1+\mathcal{O}\left(r^{2}\right)\right]+c_{2}\left(\frac{Q}{r}\right)^{2}\left[1+\mathcal{O}\left(r^{0}\right)\right],
\end{equation}
where $c_{1}$ is an arbitrary integration constant and $b_{1}$ and $c_{2}$ are known parameters whose values are shown in what follows. The value of $c_{1}$ must be estimated by numerical integration and will have some unknown dependence on $M$, $Q$, $\alpha$ and $l_{\rm P}$.  Since, in view of Fig. \ref{Fig:NakedSing}, $\psi$ approaches $\psi_{-}$ in the $r\to0$ limit, this ensures that compactness has to diverge towards negative infinity, so this implies that $c_{1}<0$. The exponent of the first term is found to be
\begin{equation}
    b_{1}=-\frac{2(1+\alpha)\left[1-\alpha+\sqrt{\alpha(\alpha-1)}\right]}{\alpha-1},
\end{equation}
whereas the second coefficient is
\begin{equation}
c_{2}=-\frac{\alpha}{2}\left\{2\sqrt{\alpha(\alpha-1)}-\alpha\left[\alpha+1-\sqrt{\alpha(\alpha-1)}\right] \right\}^{-1}.
\end{equation}
Therefore, in the $r\to0$ limit, the compactness has a divergence whose strength is modulated, in essence, by the value of the regulator. For large $\alpha$ we essentially recover the behaviour from the classical Reissner-Nordstöm geometry, as expected for a fully suppressed RSET, whereas in the $\alpha\to1$ limit, $b_{1}$ diverges. Although for $\alpha > 1$ the RSET is in itself regular, it can become divergent when evaluated over non-regular configurations. The feedback mechanism provided by the semiclassical equations reinstall a divergence at $r = 0$. An already singular geometry is being coated by a cloud of negative mass coming from vacuum polarization which nourishes from this singularity. Hence, it is expected that the leading divergence in $C$ becomes more than polynomially strong in the $\alpha\to1$ limit. In consequence, and as long as $\alpha<\frac{1}{2}\left(1+\sqrt{17}\right)$, the first term in \eqref{Eq:CompExp} dominates over the second one. For large $\alpha$, however, the electromagnetic charge carries the leading-order divergence. 

There exists an intermediate situation where both terms in \eqref{Eq:CompExp} diverge at the same rate as $r\to0$. This occurs for the exact value $\alpha=\frac{1}{2}\left(1+\sqrt{17}\right)$, for which $b_{1}=-2$.  Replacing this value of the regulator $\alpha$ in \eqref{Eq:Compasympt} and integrating yields
\begin{equation}
C=\frac{\left(1+\sqrt{17}\right)Q^{2}}{4r^{2}}\log r+\mathcal{O}\left(r^{-2}\right).
\end{equation}
This solution acts as the separatrix between two distinct behaviours of the compactness function at the singularity. It describes a situation where the divergent accumulation of vacuum polarization becomes comparable to that of the mass contribution coming from the (singular) electromagnetic source. Then, both contributions intertwine and give rise to a dominant divergence in $C$ that depends on $Q$ times a logarithmic contribution. We consider somewhat remarkable that a particular regulation of the strength of semiclassical corrections within the vicinity of the radial origin is capable to enhance divergent contributions coming from classical sources, whose sole interaction with quantum fields is through spacetime geometry.

In summary, this section has illustrated that the over-charged regime, which guarantees the presence of a naked singularity, shows a hierarchy of divergent behaviours depending on how we adjust the regulator. For small $\alpha$, the dominant contribution to the compactness comes from the backreaction of vacuum polarization on the vicinity of the singularity. However, if the RP-RSET is sufficiently dampened, the contribution coming from the singular charge-source giving rise to the electromagnetic field is uncovered. In between both regimes there exists a separatrix solution where the dominant divergence becomes a mix of quantum and classical contributions.

\section{Quasi-extremal regime}
\label{Sec:Extremal}

Always starting from an asymtoticaly flat region, we
have explored solutions where $\psi$ diverges at some radius $r>r_{\text{div}}$, and solutions where $\psi$ encounters the root $\mathcal{R}_{2}$, thus having a maximum and avoiding the formation of a minimal surface, extending the solution to $r=0$. In between both regimes there exists a separatrix solution, where $\psi$ diverges at precisely $r_{\text{div}}$, reminiscent of the classical extremal black-hole solution. In the present section we will characterize this separatrix solution.

The separatrix solution corresponds to the case in which the divergence in $\psi$ takes place at the same radius where $\mathcal{R}_{2}$ diverges, which is at $r=r_{\text{div}}$. Furthermore, this divergence in $\psi$ occurs at the same pace as $\mathcal{R}_{2}$ blows up, that is, as $(r-r_{\text{div}})^{-1}$. This can be verified by assuming the following behaviour for the $\psi$ function
\begin{equation}\label{Eq:PsiLambda}
\psi=\frac{1}{\lambda (r-r_{\text{div}})^{\gamma}}, \quad \text{with } \gamma>0.
\end{equation}
The present separatrix behaviour has the particularity that taking the limit $r\to r_{\text{div}}$ in \eqref{Eq:eqdif} reveals that the coefficient $A_{3}$ vanishes as $A_{3}\propto(r-r_{\text{div}})$. This can be checked by factorizing the numerator of $A_{3}$ \eqref{Eq:eqdif}, which is a polynomial of sixth degree for the radial coordinate, and realizing that one of its roots equals $r_{\text{div}}$ \eqref{Eq:rdiv}. In virtue of \eqref{Eq:PsiLambda} and its derivative, the following approximate relation is derived from \eqref{Eq:eqdif}:
\begin{equation}\label{Eq:ExpansionExtr}
-\frac{\gamma}{\lambda}(r-r_{\text{div}})^{-\gamma-1}=\frac{B}{\lambda^{3}}(r-r_{\text{div}})^{-3\gamma+1}+\frac{A_{2}(r_{\text{div}})}{\lambda^{2}}(r-r_{\text{div}})^{-2\gamma}+\mathcal{O}\left[(r-r_{\text{div}})\right]^{-\gamma},
\end{equation}
where $B=A_{3}/(r-r_{\text{div}})|_{r=r_{\text{div}}}$ has a finite value at $r=r_{\text{div}}$. The above expression enforces $\gamma=1$ for consistency. Therefore, both the quadratic and cubic terms in $\psi$ from Eq. \eqref{Eq:eqdif} enter the differential equation for $\psi$ at leading order in the $r\to r_{\text{div}}$ limit. Simplifying terms, the expression reduces to a quadratic equation for $\lambda$
\begin{equation}\label{Eq:PsiCoeff}
\lambda^{2}+\lambda A_{2}(r_{\text{div}})+B=0,
\end{equation}
Solutions of Eq. \eqref{Eq:PsiCoeff} return two values of $\lambda$, which we denote as $\lambda_{\pm}$ (the $\pm$ label here denotes the sign of $\lambda$). We have found the explicit analytical expressions of $\lambda_{\pm}$ by using Mathematica and they correspond to very lengthy expressions which we avoid showing here. The coefficient $\lambda_{-}$ is negative and hence comes in conflict with our integration condition from the asymptotic region. The presence of an outer horizon at $r=r_{\text{div}}$, for which the redshift function vanishes at $r=r_{\text{div}}$ with a positive slope implies $\psi$ must be positive and divergent there (when approaching the horizon from $r>r_{\text{div}}$). Hence, only the constant $\lambda_{+}$ reproduces the required behavior of an outer horizon. Therefore, in the following we choose $\lambda=\lambda_{+}$ in order to derive the approximate form of the geometry.

\begin{figure}
\label{Fig:Coefficient}
\centering
\includegraphics[width=0.8\columnwidth]{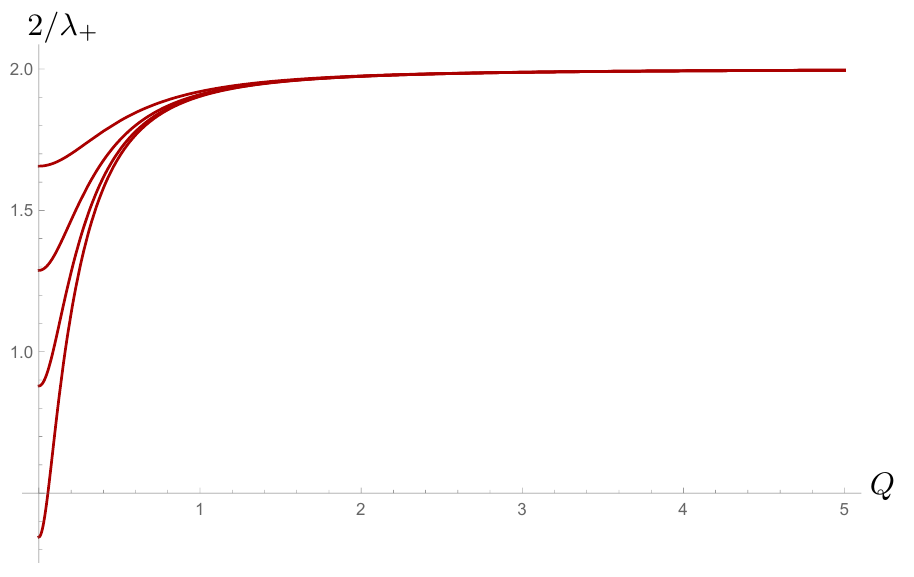}
\caption{Plot of the exponent of the redshift function in terms of the charge $Q$ for various values of $\alpha$ (from bottom to top, $\alpha$ takes the values $1.01, 2, 4$ and $10$, respectively). In the limit $Q\to0$ it goes to a constant value depending on $\alpha$ while for large $Q$ it approaches the extremal black hole solution from below. The extremal black hole is only recovered for $\abs{Q}\to\infty$. For small $Q$ the redshift function can have a Schwarzschild-like horizon.}
\end{figure}

Given that $\psi=\phi'$ by definition, Eq. \eqref{Eq:PsiLambda} with $\gamma=1$ and $\lambda=\lambda_{+}$  leads to the redshift function
\begin{equation}\label{Eq:RedsExtr}
e^{2\phi}\simeq f_{0}\left|\frac{r-r_{\text{div}}}{r_{\text{div}}}\right|^{2/\lambda_{+}},
\end{equation}
with $f_{0}$ a positive integration constant. The exponent at which the redshift function vanishes is modulated by $\lambda_{+}$, which has an involved dependency on $\alpha, l_{\rm P}$, and $Q$. For example, figure~\ref{Fig:Coefficient} shows a plot of the exponent $2/\lambda_{+}$ in terms of the charge $Q$ for various values of the regulator. 
We observe that, for $Q\gg l_{\rm P}$, it goes as 
\begin{equation}\label{Eq:lambdaclas}
\frac{2}{\lambda_{+}}= 2-\frac{4 l_{\rm P}^{2}}{Q^{2}}+\mathcal{O}\left(\frac{l_{\rm P}^{4}}{Q^{4}}\right),
\end{equation}
and the extremal Reissner-Nordström solution (for the $r>M=Q$ geometric patch) is recovered only in the limit of $l_{\rm P}=0$. 
On the other hand, the compactness function can be derived from \eqref{Eq:rrsemi}
\begin{equation}
C=1-\kappa\left(\frac{r-r_{\text{div}}}{r_{\text{div}}}\right)^{2}+\mathcal{O}\left[(r-r_{\text{div}})^{3}\right],
\end{equation}
with 
\begin{equation}
\kappa=\frac{\lambda_{+}^{2}(r_{\text{div}}^{2}-Q^{2})(r_{\text{div}}^{2}+\alpha l_{\rm P}^{2})}{l_{\rm P}^{2} r_{\text{div}}^{2}}>0.
\end{equation}
Given the classical limits of the quantities $r_{\text{div}}$ in Eq. \eqref{Eq:rdivclas} and $\lambda_{+}$ in Eq. \eqref{Eq:lambdaclas}, we obtain $\kappa=1$ for $Q\gg l_{\rm P}$, thus reproducing the extremal Reissner-Nordström compactness function.

The local form of the metric for $r>r_{\text{div}}$  in this solution has the form
\begin{equation}\label{Eq:ExtremalMetric}
ds^{2}\simeq-f_{0}\left|\frac{r-r_{\text{div}}}{r_{\text{div}}}\right|^{2/\lambda_{+}}dt^{2}+\left[\sqrt{\kappa}\left(\frac{r-r_{\text{div}}}{r_{\text{div}}}\right)\right]^{-2}dr^{2}+r^{2}d\Omega^{2}.
\end{equation}
This metric has several interesting features, which we turn to describe. One characteristic that may grab the attention is that the redshift function of the geometry goes to zero with a smaller exponent than in the classical extremal solution. The radial part of the geometry, however, retains the quadratic dependence on $r-r_{\text{div}}$; 
the affine distance between the horizon $r_{\text{div}}$ and any point of the spacetime is infinite, as in ordinary extremal spacetimes. The RP-RSET has once more caused a clear distinction between both components of the metric. This is why we refer to these solution as “quasi-extremal”.

Now we shall continue the solutions beyond $r_{\text{div}}$. We could do that in a  completely symmetric fashion. However, this would imply, on the one hand, that the radius would diminish inwards, but moreover, that the solution is no longer a proper vacuum solution as it would correspond to having a null shell localized at the horizon.
On the contrary, we can select a solution with a negatively diverging behaviour at $r_{\text{div}}$, $\psi=-1/\lambda(r_{\text{div}}-r)$, assuming now $r<r_{\text{div}}$.
This selection breaks the symmetry of the construction and is more akin to a quantum version of the extremal RN solution. Consistency with the idea that now the redshift function slope should be positive makes us to select again the $\lambda_+$ value inside the horizon. Thus, the final local form of the metric is (\ref{Eq:ExtremalMetric}), which is now valid for a sufficiently small open interval containing $r=r_{\text{div}}$.

One first implication of this solution is that, in crossing the horizon, one changes from the unconcealed branch to the concealed branch of the semiclassical corrections. Recall that the concealed branch has no well-defined classical limit. Therefore, this entire solution cannot be found perturbatively starting from the extremal RN solution. This will be more clearly seen when analyzing the next order expansion of the metric around $r=r_{\text{div}}$ (see right below). Thus, our self-consistent analyses reveal that horizons enforce non-perturbative backreaction effects, typically eliminating the horizons whatsoever, or at most maintaining an extremal-like horizon with the special characteristics we are describing.

Another notable difference with the classical extremal black-hole resides in the value of the exponent in the redshift function, $1/3 < 2/\lambda_+ < 2$ (in the $\alpha\to1$ limit). The necessary absolute value in \eqref{Eq:RedsExtr} spoils the analyticity of the metric at the “quasi-extremal” horizon $r_{\rm div}$. Despite this unusual behaviour of the metric, curvature scalars calculated from \eqref{Eq:ExtremalMetric} that are quadratic in the Ricci and Riemann tensors are finite and analytic at $r_{\rm div}$ (although a different type of singularity is indeed present, as seen below). This somewhat surprising result comes from the fact that any appearance of the function $\psi$ in the Kretschmann scalar \eqref{Eq:Kretschmann} is accompanied by $C$ factors which complete eliminate potential divergences.

The finitude and analyticity of curvature invariants at leading order comes from the particular form of \eqref{Eq:ExtremalMetric}, which describes the leading-order contributions in an expansion around $r=r_{\text{div}}$. The next order in the expansion introduces additional terms in $\psi$ which do not coincide with the classical solution in the $l_{\rm P} \to 0$ limit. As mentioned above, this is another clear indication that the “quasi-extremal” solution has an inherently quantum nature. Let us illustrate this point by assuming $\psi$ acquires an additional contribution:
\begin{equation}\label{Eq:Psipsi}
\psi=\frac{1}{\lambda_{+}(r-r_{\text{div}})}+\Psi(r),
\end{equation}
where $\Psi(r)$ is some function that diverges more slowly than $(r-r_{\text{div}})^{-1}$, or does not diverge at all. Inserting this ansatz in \eqref{Eq:eqdif}, expanding in $r-r_{\text{div}}$, and taking into account the cancellation of the leading order contributions, we obtain the expression
\begin{equation}\label{Eq:ExpansionExtr2}
\Psi'\simeq \left(A_{1}+2A_{2}\Psi+\frac{3 B \Psi}{\lambda_{+}}\right)\left[\frac{1}{\lambda_{+}(r-r_{\text{div}})}+\Psi\right]-\Psi^{2}\left[A_{2}-B\left(r-r_{\text{div}}\right)\Psi\right]+A_{0},
\end{equation}
where all coefficients are evaluated at $r=r_{\text{div}}$.

Let us assume first that $\Psi$ diverges slower than 
$1/(r-r_{\text{div}})$. This assumption implies that the leading form of the previous equation is 
\begin{equation}
\Psi' = \left(2A_{2}+\frac{3B}{\lambda_{+}}\right)\lambda_{+}^{-1} {\Psi \over (r-r_{\text{div}}) },
\end{equation}
which after solving implies 
\begin{equation}
\Psi \propto {1 \over (r-r_{\text{div}}) },
\end{equation}
but this is against our starting assumption. Therefore $\Psi$ cannot diverge at $r=r_{\text{div}}$.

Hence, the only remaining possibility for the consistency of \eqref{Eq:ExpansionExtr2} is that $\Psi$ equals a constant that causes the vanishing of all divergent contributions. An extra term linear in $r-r_{\text{div}}$ can be added to \eqref{Eq:Psipsi} so that the remaining constant terms vanish as well. The solution $\psi$ takes the form
\begin{align}\label{Eq:PsiExtremal2}
\psi
&
=\frac{1}{\lambda_{+}(r-r_{\text{div}})}+\Psi_{1}+\Psi_{2}(r-r_{\text{div}})+\mathcal{O}\left[(r-r_{\text{div}})^{2}\right],\nonumber\\
\Psi_{1}
&
=-\frac{A_{1}\lambda_{+}}{3B+2A_{2}\lambda_{+}},
\nonumber\\
\Psi_{2}
&
=\frac{3B\Psi_{1}^{2}+\lambda_{+}\left[A_{0}+\left(A_{1}+A_{2}\Psi_{1}\right)\Psi_{1}\right]}{-3B+\lambda_{+}\left(\lambda_{+}-2A_{2}\right)}\lambda_{+}.
\end{align}
The coefficients $\Psi_{1}$ and $\Psi_{2}$ fail to return the extremal black hole solution in the classical limit. The main reason behind this is linked to how semiclassical modifications expand the space of solutions. Indeed, the quasi-extremal geometry emerges as a consequence of the terms cubic in $\psi$ the RSET introduces in \eqref{Eq:eqdif} and, as a consequence, it does not connect with the extremal black hole solution when these cubic terms are suppressed. From a different perspective, as we already mentioned, part of the support of the solution lies within the concealed branch, which is intrinsically quantum and has no classical limit.

Replacing expression \eqref{Eq:PsiExtremal2} in Eq. \eqref{Eq:rrsemi} we obtain the following approximate expression for the compactness
\begin{equation}\label{Eq:CompQuasiExtr}
1-C=\kappa\left(\frac{r-r_{\text{div}}}{r_{\text{div}}}\right)^{2}\left[1+\kappa_{1}(r-r_{\text{div}})+\mathcal{O}\left(r-r_{\text{div}}\right)^{2}\right],
\end{equation}
with 
\begin{equation}
\kappa_{1}=\frac{2\lambda_{+}}{l_{\rm P}^{2}r_{\text{div}}}\left\{\lambda\left[r_{\text{div}}^{2}+l_{\rm P}^{2}\left(\alpha+r_{\text{div}}\Psi_{1}\right)\right]-\frac{l_{\rm P}^{2}Q^{2}}{r_{\text{div}}^{2}-Q^{2}}+\frac{\alpha l_{\rm P}^{4}}{r_{\text{div}}^{2}+\alpha l_{\rm P}^{2}}\right\}>0.
\end{equation}
The Kretschmann scalar \eqref{Eq:Kretschmann} takes the form
\begin{equation}\label{Eq:KretschmannDer}
\mathcal{K}=\frac{4\left(\lambda_{+}^{4}+\kappa^{2}\right)}{\left(\lambda_{+}r_{\text{div}}\right)^{4}}\left\{1+\frac{\kappa^{2}r_{\text{div}}\left(2+\lambda_{+}\right)\left(\kappa_{1}+2\lambda \Psi_{1}\right)-4\lambda_{+}^{4}}{r_{\text{div}}\left(\lambda_{+}^{4}+\kappa^{2}\right)}(r-r_{\text{div}})+\mathcal{O}\left[(r-r_{\text{div}})^{2}\right]\right\}.
\end{equation}
The constant term corresponds to the ``bulk" contribution coming from the leading-order contributions in the line element \eqref{Eq:ExtremalMetric}. Additional vanishing terms appear in curvature invariants when subdominant contributions are taken into account. The quasi-extremal metric is non-analytic at the horizon but, due to its particular form, this behavior does not reflect on curvature invariants. Similar tendencies were found in \cite{Trivedi1992, Barbachoux2002} in the context of extremal black holes in dilatonic gravity coupled to a quantum scalar field, in the sense that quantum-corrected extremal geometries develop non-analyticities at the corrected horizon. 

{We can appeal to another notion of curvature singularity that does not rely on curvature scalars. This is the definition given in \cite{Ellis1977} of non-scalar singularities: those where the components of curvature tensors, when evaluated for a suitable tetrad at the singular region, show divergences. By suitable we mean a particular tetrad field which is parallel transported along a physical curve that approaches the singular point. For this particular case, we choose a tetrad field associated to an ingoing timelike geodesic path on the metric \eqref{Eq:ExtremalMetric}. In terms of a spherically symmetric spacetime of the form \eqref{Eq:LineElement} we have
\begin{equation}\label{Eq:Ingoing}
\frac{dr}{d\tau}=-\frac{\sqrt{1-C}\sqrt{1-e^{2\phi}}}{e^{\phi}}
\end{equation}
for ingoing geodesics, where $\tau$ is a proper time which equals the coordinate time $t$ asymptotically. Replacing the components of the local metric \eqref{Eq:ExtremalMetric} in \eqref{Eq:Ingoing} we obtain
\begin{equation}\label{Eq:Geod}
\frac{dr}{d\tau}=-\sqrt{\kappa/f_{0}}\left|\frac{r-r_{\text{div}}}{r_{\text{div}}}\right|^{1-1/\lambda^{+}}.
\end{equation}
For $\lambda_{+}>1$, the velocity of the observer when crossing the quasi-extremal horizon vanishes. Moreover, Eq. \eqref{Eq:Geod} reveals that subsequent derivatives of the velocity (acceleration, jerk, and so on) are divergent for an increasing range of values of $\lambda_{+}$. The tangent vector to the ingoing radial geodesic \eqref{Eq:Geod} can be used as one of the vectors of the desired tetrad field 
\begin{equation}\label{Eq:Tetrad}
e^{\mu}_{(0)}=f_{0}^{-1}\left\{\left|\frac{r_{\text{div}}}{r-r_{\text{div}}}\right|^{2/\lambda_{+}},\sqrt{\kappa f_{0}}\left|\frac{r-r_{\text{div}}}{r_{\text{div}}}\right|^{1-1/\lambda_{+}},0,0\right\}.
\end{equation}
To prove there is a non-scalar curvature singularity at $r=r_{\text{div}}$, it is sufficient to obtain a single divergent component of the Riemann tensor contracted with the vector field \eqref{Eq:Tetrad} of the tetrad. For simplicity, we select the component
\begin{equation}
R_{(0)\theta\theta(0)}=R_{t\theta\theta t}e^{t}_{(0)}e^{t}_{(0)}+R_{r\theta\theta r}e^{r}_{(0)}e^{r}_{(0)}=\frac{\kappa}{f_{0}}\left(\frac{\lambda_{+}-1}{\lambda_{+}}\right)\left|\frac{r-r_{\text{div}}}{r_{\text{div}}}\right|^{1-2/\lambda_{+}}.
\end{equation}
Note that, for $1<2/\lambda_{+}<2$, this physical component of the curvature is singular. We only recover a well-defined geometry in the classical limit $\lambda_{+}\to1$. Note that the curvature singularity at the quasi-extremal horizon follows from the non-analyticity of the metric for $\lambda_{+}>1$.
If  the exponent of the redshift function takes values below $1$, divergences will appear at higher-order derivatives of the components of the Riemann curvature tensor. We have thus proved that the quasi-extremal black hole has a non-scalar curvature singularity at the horizon.

The extremal black hole is more stable against quantum perturbations than black holes with an outer horizon. As we have seen, vacuum polarization eliminates these horizons replacing them with a wormhole neck. The particular case of the extremal black hole results in a different kind of modification, in the sense that the horizon itself is preserved and becomes a curvature singularity. An important difference between these two cases can be understood by looking at the geometry beyond the quasi-extremal horizon. The shape of the solution can be again univocally determined by arguments concerning the roots and exact solutions. Figure \ref{Fig:Extremal} contains a plot of one of these solutions, showing details of the inner part of the geometry. In the inner region $r<r_{\text{div}}$, the solution $\psi$, now living in the concealed branch, approaches $-\infty$ from below the root $\mathcal{R}_{2}$. Then, it crosses $r=Q$ in a similar fashion as it occurred in the over-charged regime (see Sec. \ref{Sec:Naked}), with the difference that the exact solution intersected is now $\psi_{+}$. After intersecting the exact solution, it meets the condition for a branch jump given by the vanishing of $\mathcal{G}$ \eqref{Eq:Gcoeff} and the solution switches  to the unconcealed branch back again. More explicitly, the branch jump requires that the solution $\psi$ takes the value
\begin{equation}
\psi(r_{\text{jump}})=-\frac{r_{\text{jump}}^{2}+\alpha l_{\rm P}^{2}}{l_{\rm P}^{2}r_{\text{jump}}}.
\end{equation}

Once the transition to the unconcealed branch has occurred, the solution stays within this branch, growing until it reaches a maximum and remaining trapped between $\mathcal{R}_{2}$ and $\psi_{-}$. The behaviour of $\psi$ close to the radial origin is the same as in the over-charged regime. The main difference with the over charged regime is that the central singularity is no longer naked, but covered by a likewise singular horizon at the end of an infinite wormhole neck. It is interesting to recall that this is the separatrix between two solutions both harboring singularities not covered by event horizons of any sort. These can be located either at radial infinity in the wormhole regime, or at the radial origin for the over-charged family. Beyond its neck, the wormhole solution stays within the concealed branch, extending towards an asymptotically singular region at radial infinity. However, since the quasi-extremal solution does not have its radial coordinate reverted but just elongated through an infinite tube, it extends towards $r=0$. Therefore, below the (singular) quasi-extremal horizon there is just a narrow region where the solution is non-perturbative, and the perturbative regime is recovered once we go sufficiently deep beyond the horizon. Nevertheless, this narrow band where quantum corrections become non-perturbative, as well as the non-scalar singularity, persist under changes of the regulator, only disappearing in the limit of infinite charge. Thus, we also expects that for soluctions with $M=Q_{\rm crit} >> l_{\rm P}$, the main characteristics of the quasi-extremal solution described in here will be preserved in more refined approximations to the RSET.
\begin{figure}
\centering
\includegraphics[width=0.8\columnwidth]{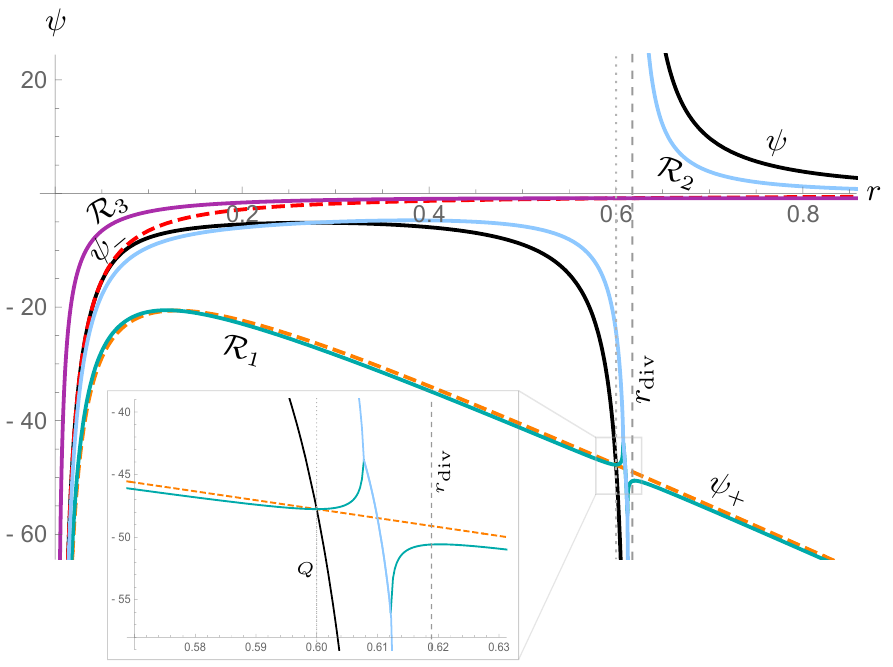}
\caption{Numerical plot of a quasi-extremal solution (black curve), with $Q=0.60$, $M\simeq0.59$ and $r_{\text{div}}\simeq0.62$. The roots and exact solutions appear represented, together with the vertical dashed and dotted lines denoting $r=r_{\text{div}}$ and $r=Q$, respectively. The solution has a branch jump at the quasi-extremal horizon. In the region $r<r_{\text{div}}$, $\psi$ intersects the exact solution $\psi_{+}$ and the root $\mathcal{R}_{1}$ exactly at $r=Q$ (zoomed figure). Then, the solution jumps to the unconcealed branch at $r_{\text{jump}}\simeq0.58$ and grows towards smaller $r$ until crossing $\mathcal{R}_{2}$, reaching a maximum, and starts decreasing confined between $\mathcal{R}_{2}$ and $\psi_{-}$. The asymptotic behaviour near $r\simeq0$ is the same as for super-charged solutions.}
\label{Fig:Extremal}
\end{figure}
%

\section{Final remarks}
\label{Sec:Conclusions}

In this work we have obtained the complete set of self-consistent electro-vacuum solutions in the semiclassical approximation taking the RP-RSET as describing the quantum material content of the spacetime. For the Boulware vacuum state, the only state compatible with staticity and asymptotic flatness, three families of solutions have been obtained depending on the charge-to-mass ratio.

Under-charged solutions ($Q<Q_{\text{crit}}$) are wormhole geometries with essentially the same features as the Schwarzschild geometry counterpart from \cite{Arrechea2019}. An increase of charge exerts a repulsive effect that makes the wormhole neck shrink, which nevertheless always sits above the classical gravitational radius $r_{+}$ \eqref{Eq:RNHor}. On the other side of the neck, there is a null naked singularity at finite affine distance for all geodesics. The regulator $\alpha$ allows to extend the space of solutions to wormholes of Planckian size, something forbidden for the Polyakov RSET due to its unphysical singularity at $r=l_{\rm P}$. In any case, in physical terms solutions with $M \sim l_{\rm P}$ should not be very trustable as the semiclassical approximation should break down in that regime.

Over-charged solutions ($Q>Q_{\text{crit}}$) describe naked singularities at $r=0$. This can be thought of as composed by a cloud of infinite negative mass coming from vacuum polarization, originated from the backreaction of vacuum energy caused by the infinite charge density of the electromagnetic field. Depending on the characteristics of the regulator, the contribution of this cloud of vacuum polarization can be suppressed so that the dominant singular contribution at short distances comes from the electromagnetic SET. This geometry exemplifies a situation where slight modifications in the regulator parameter $\alpha$ strongly change the features of the solution close to $r=0$. 

Lastly, the appeareance of a “quasi-extremal” geometry ($Q = Q_{\rm crit}$) partially responds to a question raised in our previous work \cite{Arrechea2019}, where we wondered whether semiclassical gravity would allow for the existence of horizons where the roots of the redshift function have a greater multiplicity. It can be arguably expected that the Boulware state would be able to coexist with horizons of the extremal kind, as these represent zero-temperature configurations. The quasi-extremal case here studied exemplifies this situation. The backreacted geometry that we find is more alike to its classical counterpart than geometries of the wormhole kind, with the caveat that non-perturbative corrections still occurr in a narrow region behind the singular horizon. We find that indeed quantum backreaction retains the horizon structure; however, it transforms it from having a parabolic shape to having a cusp-like shape, its sharpness being modulated by the exponent $\lambda^+$. Then, this cusp translates into a non-scalar curvature singularity of the kind defined in  \cite{Ellis1977}.

The incompatibility between Schwarzschild-like horizons (at nonzero temperature) and the Boulware state can be interpreted as an indication that trapping horizons must be dynamical and be subjected to an evaporation process (see \cite{Arrechea2020} for a discussion on the possible outcomes of dynamical evaporation scenarios). The case of the extremal horizon is exceptional since, due to having zero temperature, in principle they do not need entering into an evaporation regime. It is reasonable to interpret our non-scalar curvature singularity result as pointing out that extremal configurations have additional elements of non-physicality beyond those already present at the classical level.

In view of our analyses, we conclude that semiclassical electro-vacuum static geometries in the Boulware vacuum state show the unavoidable appearance of curvature singularities, whose spacetime location depends on a balance between charge and mass. Semiclassical self-consistency ensures that these singular geometries are devoid of event horizons of any kind. The sole exception is the ``quasi-extremal" solution, for which the horizon itself constitutes a singularity. These results indicate the incompatibility between the presence of quantized fields in the Boulware state and regular event horizons. 

\acknowledgments

Financial support was provided by the Spanish Government through the projects FIS2017-86497-C2-1-P, FIS2017-86497-C2-2-P (with FEDER contribution), FIS2016-78859-P\linebreak (AEI/FEDER,UE), and by the Junta de Andalucía through the project FQM219. Authors JA and CB acknowledge financial support from the State Agency for Research of the Spanish MCIU through the
``Center of Excellence Severo Ochoa" award to the Instituto de Astrofísica de Andalucía (SEV-2017-0709).

\bibliographystyle{unsrt}
\bibliography{biblio3}
\end{document}